\documentclass[aps,prl,twocolumn,groupedaddress,amsmath]{revtex4}

\usepackage{graphicx}

\begin{document}

\title{Simulating rare events in equilibrium or non-equilibrium stochastic systems}

\author{Rosalind J. Allen}

\email{R.Allen@amolf.nl}

\affiliation{FOM Institute for Atomic and Molecular Physics, Kruislaan 407, 1098 SJ Amsterdam, The Netherlands}

\author{Daan Frenkel}

\affiliation{FOM Institute for Atomic and Molecular Physics, Kruislaan 407, 1098 SJ Amsterdam, The Netherlands}

\author{Pieter Rein ten Wolde}

\affiliation{FOM Institute for Atomic and Molecular Physics, Kruislaan 407, 1098 SJ Amsterdam, The Netherlands}

\date{\today}

\begin{abstract}
We present three algorithms for calculating rate constants and sampling transition paths for rare events in simulations with stochastic dynamics. The methods do not require {\em{a priori}} knowledge of the phase space density and are suitable for equilibrium or non-equilibrium systems in stationary state. All the methods use a series of interfaces in phase space, between the initial and final states, to generate transition paths as chains of connected partial paths, in a ratchet-like manner. No assumptions are made about the distribution of paths at the interfaces. The three methods differ in the way that the  transition path ensemble is generated. We apply the algorithms to kinetic Monte Carlo simulations of a genetic switch and to Langevin Dynamics simulations of intermittently driven polymer translocation through a pore. We find that the three methods are all of comparable efficiency, and that all the methods are much more efficient than brute force simulation.
\end{abstract}

\pacs{}

\maketitle

\section{Introduction}
Rare events are fluctuation-driven processes which occur infrequently. Many natural processes can be classified as rare events, including the nucleation of crystals or protein aggregates, chemical reactions, earthquakes and some meteorological phenomena.  For these processes, the average waiting time between events is orders of magnitude longer than the timescale of the event itself. In this situation, conventional  ``brute force'' simulation is highly inefficient.  This is because few, if any, events are likely to happen in the accessible simulation time, and the  majority of the computational effort is spent in simulating the uneventful waiting time. Simulation of rare events requires the use of specialized techniques, such as Bennett-Chandler methods \cite{daan,chandler} or Transition Path Sampling \cite{dellago,tps,bolhuis}. Such techniques have been extensively used for problems including crystal nucleation, membrane permeation, ion transfer reactions and peptide folding. However, these methods require knowledge of the phase space density in the initial state and as a result they are  only suitable for (possibly metastable) equilibrium systems. By ``equilibrium'', we mean systems where detailed balance is satisfied and the phase space density is known. For example, in equilibrium at constant particle number, volume and temperature (NVT), the phase space density follows the Boltzmann distribution. For non-equilibrium systems in steady state - {\em{i.e.}} systems in which there are, on average, probability currents in phase space - the steady-state phase space density is generally not known {\em{a priori}}. Consequently, ``conventional'' rare event techniques cannot be used for non-equilibrium systems. In this paper, we present several techniques that do not require knowledge of the phase space density and are therefore suitable for rare events in steady-state systems in or out of equilibrium.

Rare events in non-equilibrium systems constitute a host of important problems that have thus far been generally inaccessible to simulations. Examples include crystal nucleation under shear, polymer conformational transitions in hydrodynamic flows, driven transport through membranes and most rare events in biological systems. To our knowledge, the only scheme to have been proposed for obtaining transition paths for rare events out of equilibrium in stochastic dynamical systems is that of Crooks and Chandler \cite{crooks}. Here, transition trajectories (paths) connecting the initial and final states are characterized by their random number history. New transition paths are generated by making changes in the random number history of previously generated paths. This method requires that paths do not diverge significantly upon changing the random number history; for high dimensional systems, the Lyapunov instability is likely to lead to inefficiency.

For equilibrium systems, a variety of rare event techniques exist. Some of these, for example  Bennet-Chandler methods \cite{daan,chandler}, involve the calculation of the free energy along a pre-determined reaction co-ordinate. These methods do not generate transition trajectories and moreover,  choice of an inappropriate co-ordinate leads to inefficient calculation of the rate constant. Other methods, such as  Transition Path Sampling \cite{dellago,tps,bolhuis} and  Transition Interface Sampling \cite{vanerp,vanerp2005}, do not require the specification of a reaction co-ordinate and do generate transition paths. These methods require that the transition occurs very rapidly, since new paths are generated by a shooting procedure and tend to diverge, leading to inefficiency. String methods  \cite{e2002,e2005} have also been developed, but have not yet been implemented for large systems. Finally, several methods, such as Milestoning \cite{faradjian} and Partial Path Transition Interface Sampling \cite{moroni} use a series of interfaces in phase space, like the methods to be discussed here. However, these techniques assume that the distribution of transition paths at the interfaces follows the equilibrium distribution: an assumption which is unlikely to be justified in many cases \cite{tenwolde}.

In this paper, we discuss several alternative schemes  for calculating rates and obtaining transition paths in stochastic dynamical systems. As well as enabling the efficient simulation of rare events in non-equilibrium steady-state systems, the methods also avoid many of the difficulties associated with existing equilibrium rare event methods. The methods do not require the specification of a reaction co-ordinate, and transition paths are generated without any requirement on their length (since paths are generated by a ratchet-like procedure and not by shooting from previous paths). Furthermore, although a series of interfaces in phase space is used, no assumptions are made about the distribution of paths at the interfaces. The first method, Forward Flux Sampling (FFS), was presented in an earlier publication \cite{allen}.

 After an introduction to the theory, we give a detailed description of FFS (Section \ref{sec_ffs}), and also of  two more path sampling schemes, the ``Branched Growth'' method (Section \ref{sec_bg}) and the ``Rosenbluth'' method (Section \ref{ros_samp}). The latter methods have been developed in analogy to efficient schemes  for sampling polymer chain configurations.  The BG method also resembles a technique used for computing rare event probabilities in the field of telecommunications \cite{altamirano}. In Section \ref{sec_prune}, we discuss a ``pruning'' method for increasing the efficiency of the path sampling schemes. All three schemes are then demonstrated for two very different systems: in Section \ref{gensw}, the flipping of a genetic switch is modelled using  kinetic Monte Carlo simulations and in Section \ref{poltr}, we apply the methods to Langevin Dynamics simulations of driven polymer translocation through a pore. We discuss the methods in the context of other rare event techniques, and assess their  advantages and disadvantages in Section \ref{dis}. Finally, Appendices \ref{theor_j}, \ref{was} and \ref{prune} contain theoretical justifications of the algorithms, an alternative reweighting scheme for the Rosenbluth method and a detailed discussion of the pruning scheme. 

\section{Theoretical Background}\label{sec_rate}
\begin{figure}[h]
\begin{center}
{\rotatebox{0}{{\includegraphics[scale=0.45,clip=true]{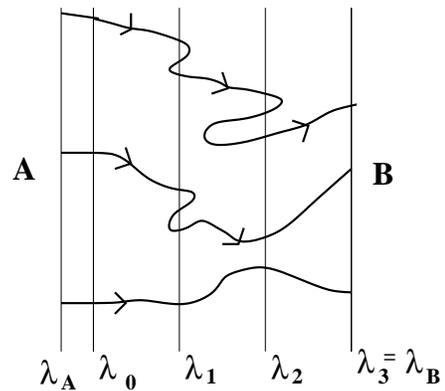}}}}
\caption{Schematic illustration of the definition of regions $A$ and $B$ and the interfaces  $\lambda_0 \dots \lambda_n$ (Here, $n=3$). Three transition paths are shown. \label{fig1_1} }
\end{center}
\end{figure}

We assume that the rare event can be viewed as a spontaneous transition between two well-defined regions of phase space $A$ and $B$; by ``phase space'', we mean the set of all parameters that characterize the system. We are interested in calculating the rate constant $k_{AB}$: the average rate of transitions from $A$ to $B$. We  use the ``effective positive flux'' expression described by Van Erp {\em{et al}} \cite{vanerp,moroni,vanerp2005}. $A$ and $B$ are defined in terms of a parameter $\lambda (x)$, such that $\lambda < \lambda_A$ in $A$ and $\lambda > \lambda_B$ in $B$. Here, $x$ denotes the co-ordinates of the phase space. A series of non-intersecting surfaces in phase space $\lambda_0 \dots \lambda_n$ are chosen, such that $\lambda_0 \ge \lambda_A$, $\lambda_n=\lambda_B$ and $\lambda_i > \lambda_{i-1}$. These must be chosen such that any path from $A$ to $B$ passes through each surface in turn, not reaching $\lambda_{i+1}$ before it has crossed $\lambda_i$. This is illustrated in Figure \ref{fig1_1}. Please note the change in notation in the numbering of the interfaces, compared to our earlier paper \cite{allen}.

 Defining the history-dependent functions  $h_{\mathcal{A}}$ and  $h_{\mathcal{B}}$ such that $h_{\mathcal{A}}=1$ and $h_{\mathcal{B}}=0$ if the system was more recently in $A$ than in $B$, and $h_{\mathcal{A}}=0$ and $h_{\mathcal{B}}=1$ otherwise, the  rate constant $k_{AB}$ for transitions from $A$ to $B$ is given by \cite{vanerp,vanerp2005}:
\begin{equation}\label{eq1}
k_{AB} = \frac{{\overline{\Phi}}_{A,n}}{{\overline{h}}_{\mathcal{A}}} = \frac{{\overline{\Phi}}_{A,0}}{{\overline{h}}_{\mathcal{A}}} P(\lambda_n|\lambda_0) .
\end{equation}
Here, ${\Phi}_{A,j}$ is the flux of trajectories with $h_{\mathcal{A}}=1$ ({\em{i.e.}} coming from $A$), that cross $\lambda_j$ for the first time: thus ${\Phi}_{A,n}$ is the flux of trajectories reaching $B$ from $A$ and ${\Phi}_{A,0}$ is the flux reaching the first interface $\lambda_0$ from $A$. The overbar denotes a time average, and the factor ${{\overline{h}}_{\mathcal{A}}}$ is the average fraction of the time that the system spends in the ``basin of attraction'' of $A$.  $P(\lambda_n|\lambda_0)$ is the probability that a trajectory that reaches $\lambda_0$ subsequently arrives in $B$ instead of returning to $A$.  Eq.(\ref{eq1}) states that  the total flux from $A$ to $B$ is the flux of trajectories from $A$ to $\lambda_0$, multiplied by the probability that such a trajectory  will later reach $B$. In this way, the problem of calculating the very small flux ${\Phi}_{A,n}$ is reduced to a calculation of a larger flux ${\overline{\Phi}}_{A,0}$, and a small probability $P(\lambda_n|\lambda_0)$. $P(\lambda_n|\lambda_0)$  can then be expressed as the product of the probabilities $P(\lambda_{i+1}|\lambda_{i})$ that a trajectory that comes from $A$ and crosses $\lambda_{i}$ for the first time, will subsequently reach $\lambda_{i+1}$ instead of returning to $A$:
\begin{equation}\label{eq2}
P(\lambda_n|\lambda_0) = \prod_{i=0}^{n-1}P(\lambda_{i+1}|\lambda_{i})
\end{equation}
It is important to point out that Eq.(\ref{eq2}) does not imply an assumption that the system is Markovian. This is because the conditional probabilities $P(\lambda_{i+1}|\lambda_{i})$ are implicitly weighted over the ensemble of paths reaching $\lambda_i$ from $A$, as shown in Appendix \ref{app_prob}. The discussion in Appendix \ref{app_prob} also shows the equivalence of the averaging procedures used to evaluate $P(\lambda_n|\lambda_0)$ in the  three path sampling methods described in this paper.

The methods described in this paper allow one to sample the Transition Path Ensemble (TPE), as well as calculating the rate constant. The TPE is the collection of all transition trajectories (paths) from $A$ to $B$ that would be obtained if an infinitely long brute force simulation were to be performed. Analysis of the TPE can lead to a mechanistic understanding of the rare event in question through, for example, the calculation of committor distributions \cite{tps}. We shall see in Sections \ref{sec_ffs}, \ref{sec_bg} and \ref{ros_samp}, as well as Appendix \ref{theor_j}, that the three methods discussed here generate transition paths belonging to the TPE in an efficient manner and with the correct weights.

The starting point for all  three methods is the choice of the parameter  $\lambda(x)$ and the definition of phase space regions $A$ and $B$. $\lambda(x)$ must increase monotonically as the interfaces $\lambda_0 \dots \lambda_n$ are crossed. However, there is no assumption that $\lambda$ is the reaction co-ordinate: transition paths are free to follow any possible path between $A$ and $B$, including paths which ``loop back'', crossing some interfaces several times. A good choice of $\lambda$ will improve the efficiency of the calculation but will not affect the final rate constant or transition paths. In fact, for the test systems of Sections \ref{gensw} and \ref{poltr}, our choice of  $\lambda$ is very simple and is unlikely to correspond to the true reaction co-ordinate.  The interfaces $\lambda_0 \dots \lambda_n$ are placed between $A$ and $B$, with $\lambda_n=\lambda_B$.  We find that it is often convenient (although not necessary) to place $\lambda_0$ at the border of the $A$ region: $\lambda_0=\lambda_A$. 
The optimum number and placement of  the interfaces will be discussed in detail in a future publication \cite{long2}.

\section{Algorithms}

\subsection{The Forward Flux Method}\label{sec_ffs}
\begin{figure}[h]
\begin{center}
{\rotatebox{0}{{\includegraphics[scale=0.65,clip=true]{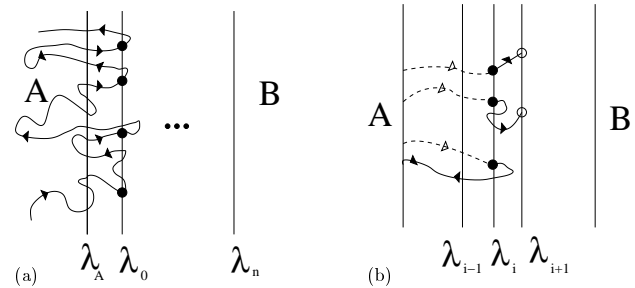}}}}
\caption{The first (a) and second (b) stages of the FFS method. The distribution of points at the interfaces depends on the history of the paths, as illustrated by the dashed lines in (b). The circles are members of the collection of points at the interfaces $\lambda_i$. \label{fig1} }
\end{center}
\end{figure}

The first of our three methods is Forward Flux Sampling (FFS), which was introduced in an  earlier paper \cite{allen}. For clarity, we describe the method again here, together with  some recent improvements. The FFS algorithm begins with a simulation run in the basin of attraction of region $A$. The parameter $\lambda$ is monitored during this run. Each time the trajectory leaves $A$ and reaches $\lambda_0$ {\em{for the first time since leaving A}}, a counter  $q$ is incremented and the phase space co-ordinates of the system are stored. The run is then continued until $N_0$ points at $\lambda_0$ have been collected, as illustrated in Figure \ref{fig1}a. If, during this run, the system happens to enter $B$, it is replaced in $A$ and re-equilibrated. The number $N_0$ of collected points should be as large as possible, in order to obtain good sampling of the transition paths. As discussed in Section \ref{dis}, if $N_0$ is so small that it is similar to the number $n$ of interfaces, sampling problems will occur due to ``genetic drift''. In the examples of Sections \ref{gensw} and \ref{poltr}, $N_0$ is of the order of thousands. The flux  ${\overline{\Phi}}_{A,0}/{\overline{h}}_{\mathcal{A}}$ is given by  ${\overline{\Phi}}_{A,0}/{\overline{h}}_{\mathcal{A}}= q/T$, where $T$ is the total length of the simulation run.  Figure \ref{fig1}a shows this first stage of the algorithm: crossings of $\lambda_0$ that are labeled with a black circle contribute to $q$ and to the collection of points at $\lambda_0$. In practice, it may be convenient not to store the co-ordinates of every ``black circle'', but rather to select crossings in some unbiased way. We note that $q$ must be incremented for every ``black circle'' crossing. 

In the next stage of the algorithm, we estimate the probability  $P(\lambda_{i+1}|\lambda_i)$ of reaching $\lambda_{i+1}$ from $\lambda_{i}$, instead of returning to $A$. Starting with the collection of $N_0$ points at $\lambda_0$, $M_0$ trial runs are carried out. Each trial run consists of selecting a point from the collection at random and using it as the starting point for a simulation run which is continued until either $\lambda_1$ or $\lambda_A$ is reached.  If the trial run reaches $\lambda_1$ then a counter $N_s^{(0)}$ is incremented, and the final point of the run is stored in a new collection of points at $\lambda_1$. After the $M_0$ trial runs, we are left with  an estimate for $P(\lambda_1|\lambda_0) = N_s^{(0)}/M_0$, 
and a collection of $N_1 = N_s^{(0)}$ points at $\lambda_1$. Using this collection of points, $M_1$ trial runs are then carried out, each time selecting a starting point at random and running a simulation until either $\lambda_2$ or $\lambda_A$ is reached. This procedure is repeated for each interface $\lambda_i$, each time using the collection of points generated at the previous interface and firing trial runs as far as $\lambda_{i+1}$ or $\lambda_A$. A possible way of improving efficiency by eliminating long paths back to $A$ will be discussed in section \ref{sec_prune} and Appendix \ref{prune}. The end result of the trial run procedure is an estimated value of $P(\lambda_{i+1}|\lambda_i) = N_s^{(i)}/M_i$, for each interface $i$. Multiplying these together as in Eq.(\ref{eq2}) leads to an estimate for $P(\lambda_n|\lambda_0)$. This can then be multiplied by the flux ${\overline{\Phi}}_{A,0}/{\overline{h}}_{\mathcal{A}}$ calculated in the first stage, to give the rate constant $k_{AB}$.

The method described here is slightly different from that outlined in our earlier paper \cite{allen}. The collection of points at $\lambda_{i+1}$ now consists of the end points of all the $N_s^{(i)}$ successful trajectories from $\lambda_i$; previously, only a user-defined number of points was stored and the rest were discarded. Storing a larger number of points at the interfaces leads to better sampling, with no increase in cost. In addition, the only user-defined parameters are now the number $N_0$ of initial points and the numbers $M_i$ of firing runs to be carried out at each interface (as well as the number and position of the interfaces).  It is important to ensure that the $M_i$ are large enough to generate sufficient points at the next interface for good sampling. In our earlier paper, we also described a procedure whereby a series of ``sub-interfaces'' between each pair of interface $\lambda_i$ and $\lambda_{i+1}$ were used to construct histograms for $P(\lambda|\lambda_i)$, which were then fitted together to obtain $P(\lambda_n|\lambda_0)$. The aim of this approach was to reduce the accumulation of errors caused by the multiplication of conditional probabilities in Eq.(\ref{eq2}). While this fitting procedure gives the correct rate constant, we find that in practice it does not improve the efficiency of the method and we do not therefore use it.

\begin{figure}[h]
\begin{center}
{\rotatebox{0}{{\includegraphics[scale=0.45,clip=true]{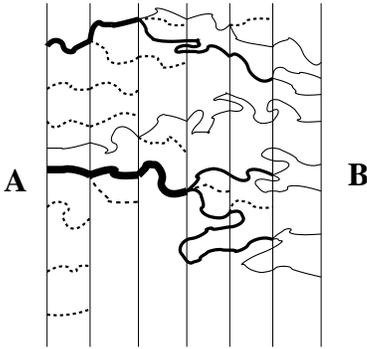}}}}
\caption{ Schematic illustration of the extraction of the transition path ensemble from the FFS procedure. All partial paths that reach the subsequent interface are shown. Partial paths that do not contribute to the TPE are shown by dotted lines. The solid lines correspond to the TPE; the width of the line indicates the weight of the contribution of a particular partial path to the TPE. \label{fig1_2} }
\end{center}
\end{figure}

The FFS method generates transition paths according to their correct weights in the Transition Path Ensemble (TPE), as shown in Appendix \ref{ffs_app}. In order to extract these paths, the phase space co-ordinates of the system must be stored for all points along all trial runs which successfully reach $\lambda_{i+1}$ from $\lambda_i$.  One must also store information on the connectivity of the partial paths; {\em{i.e.}} each successful trial from $\lambda_i$ to $\lambda_{i+1}$ is annotated with an index that describes its  initial point at $\lambda_i$. Once the trial run procedure is complete, transition paths are obtained by following the trials that reach $B$ back to $\lambda_{n-1}$, following their initial points back to $\lambda_{n-2}$, and so on back to $A$.  As illustrated in Figure \ref{fig1_2}, this results in a ``branching tree'' of transition paths, in which partial paths close to $A$ may be shared by many members of the TPE. The resolution in phase space of the TPE is therefore better for phase space regions close to $B$ than for those close to $A$; the TPE that is produced is nevertheless correctly weighted. 

 A method similar to FFS has recently been used to study the crystal nucleation of sodium chloride \cite{valeriani}.

\subsection{The Branched Growth Method}\label{sec_bg}

We now describe an alternative path sampling and rate constant calculation scheme: the ``Branched Growth'' (BG) method. Both the BG method and the Rosenbluth scheme of Section \ref{ros_samp} are similar to techniques originally developed for the efficient sampling of polymer configurations \cite{daan}; an analogy is used between  transition paths and conformations of a polymer chain, with partial paths between interfaces playing the role of  polymer segments. The BG method also resembles a technique that is used to compute probabilities of rare events in telecommunication systems \cite{altamirano}.

\begin{figure}[h]
\begin{center}
{\rotatebox{0}{{\includegraphics[scale=0.45,clip=true]{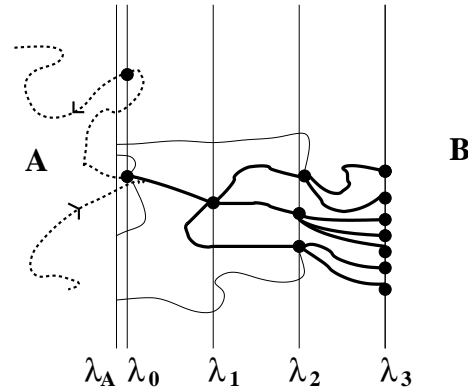}}}}
\caption{A schematic view of the generation of a branched path (thick lines) using the Branched Growth Sampling method. The simulation run in the $A$ basin is shown by a dotted line. Trial runs which fail to reach $\lambda_{i+1}$ are shown by thin lines. The generation of the initial point for the next path is also shown.  \label{fig_bg} }
\end{center}
\end{figure}

The BG method begins with a  simulation in the basin of attraction of $A$, which is suspended when the system leaves $A$ and crosses $\lambda_0$.  The resulting system configuration at $\lambda_0$ is then used as the starting point for  $k_0$ trial runs, which are continued until either $\lambda_1$ or $\lambda_A$ is reached. Each trial run is assigned a ``weight''  $1/k_0$. If $N_s^{(0)}>0$ of the trials reach $\lambda_1$, then each of the  $N_s^{(0)}$ end points at $\lambda_1$ becomes a starting point for $k_1$ new trial runs, which have weight $1/(k_0k_1)$, and which are continued until $\lambda_2$ or $\lambda_A$ is reached.  Each of the $N_s^{(1)}$ total successful trials from $\lambda_1$ to $\lambda_2$ generates a starting point for $k_2$ trial runs to $\lambda_3$, with weight $1/(k_0k_1k_2)$, and so on until $\lambda_n=\lambda_B$ is reached. Once the generation of one branching path is over, either because $B$ was reached, or because no successful trials were generated at some intermediate interface $\lambda_i$, we obtain an estimate of $P(\lambda_n|\lambda_0)$ from the total weight of the branches that eventually reaches $\lambda_n$: $P(\lambda_n|\lambda_0) = N_s^{(n-1)}/\prod_{i=0}^{n-1} k_i$. In order to begin the generation of the next branching path,  the simulation run in the  $A$ basin is resumed and a  new starting point at $\lambda_0$ is generated the next time the system crosses $\lambda_0$, coming from $A$ (the system must return to $A$ between subsequent starting point generations). The same trial run procedure is then used to create a ``branching tree'' of paths from this starting point, resulting in a new estimate of $P(\lambda_n|\lambda_0)$. After many such branching paths have been generated, the  final estimate of $P(\lambda_n|\lambda_0)$ is simply the average of the contributions due to all the paths  (including those which failed to reach $B$: these make a zero contribution). The flux ${\overline{\Phi}}_{A,0}/{\overline{h}}_{\mathcal{A}}$ can be obtained from the total number of crossings observed in the simulation in the $A$ basin, divided by the total length of this run. The ``branching trees'' of paths connecting regions $A$ and $B$ which arise from this sampling method are correctly weighted members of the TPE, as shown in Appendix \ref{ffs_app}. By storing the phase space co-ordinates of the points in the trial runs which successfully reach $\lambda_{i+1}$, one can obtain branching transition paths, which  can be analyzed to obtain information on the mechanism by which the rare event occurs.

The Branched Growth method is illustrated schematically in Figure \ref{fig_bg}. As in FFS, the TPE is sampled with better resolution in the phase space region close to $B$, since the transition paths are branched. 

\subsection{The Rosenbluth Method}\label{ros_samp}
Our third scheme, the ``Rosenbluth'' (RB) method, is related to the Rosenbluth scheme for sampling polymer chain conformations \cite{daan,rosenbluth,fms}. As for the BG method, transition paths are generated one at a time. In contrast to BG, however, the RB method generates unbranched paths. This means that the TPE is sampled evenly for all values of $\lambda$ and also makes the extraction and analysis of transition paths very easy. Furthermore, the RB method requires less storage of system configurations, which may be useful for large systems. 

 For the FFS and BG methods, we show in Appendix \ref{theor_j} that the TPE is automatically generated with the correct path weights.  However, as we shall see, in the RB method, when paths are initially generated they do not have the correct weights. A ``reweighting'' procedure is therefore needed in order to correctly sample the TPE. Here, we describe  a Metropolis-type acceptance/rejection reweighting procedure \cite{daan}; in Appendix \ref{was}, we also discuss an alternative ``Waste-Recycling'' scheme based on the recent work of Frenkel \cite{frenkel}.

\begin{figure}[h]
\begin{center}
{\rotatebox{0}{{\includegraphics[scale=0.45,clip=true]{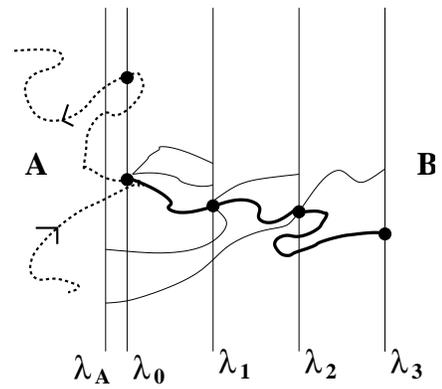}}}}
\caption{ A schematic view of the generation of a transition path using the Rosenbluth Sampling method. The simulation run in the $A$ basin is shown by a dotted line. The transition path is shown by bold lines. Trial runs which do not form part of the transition path are shown by thin lines. The generation of the next starting point at $\lambda_0$ is also illustrated. \label{fig_ros} }
\end{center}
\end{figure}

The generation of a transition path in the RB method takes place as illustrated in Figure \ref{fig_ros}. We begin with an initial point at $\lambda_0$, which is obtained in the same way as for the FFS and BG methods, using a simulation run in the basin of attraction of $A$. The point at $\lambda_0$ is used to initiate $k_0$ trial runs, which are  continued until they reach either $\lambda_1$ or $\lambda_A$. If at least one of these successfully reaches $\lambda_1$, we choose one successful trial at random, and use its end point at $\lambda_1$ as the starting point for a set of $k_1$ trial runs, which end either at $\lambda_2$ or at $\lambda_A$. Once again, a successful trial is chosen at random and used to continue the path. This procedure is repeated until either $B$ is reached, or no successful trials are produced. 

\begin{figure}[h]
\begin{center}
{\rotatebox{0}{{\includegraphics[scale=0.45,clip=true]{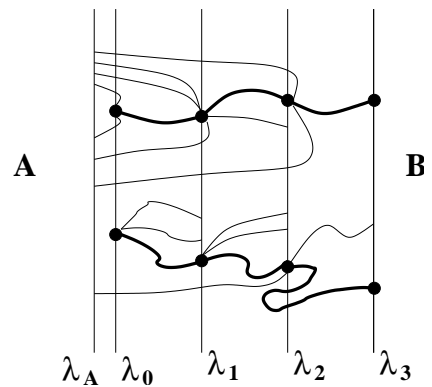}}}}
\caption{Two transition paths generated by the Rosenbluth method. The bottom path must be reweighted by a factor of 9 relative to the top path.
\label{fig_ros2} }
\end{center}
\end{figure}

The RB method generates unbranched transition paths, in contrast to the FFS and BG methods. For FFS and BG, paths for which more trial runs are successful produce more branches and make a greater contribution to the TPE, resulting in ``automatic'' correct weighting of the transition paths. In the RB method, however, one successful trial at each interface is chosen, regardless of how many successful trials there were. This leads to paths being generated with incorrect weights: as illustrated schematically in Figure \ref{fig_ros2}, paths for which more trials were successful must be given an increased weight in the TPE relative to those for which fewer trials were successful. We show in Appendix  \ref{m2_app} that the weight of each generated transition path must be  multiplied by the ``Rosenbluth factor'' $W$, which is given by:
\begin{equation}\label{wt1}
W = \prod_{j=0}^{n-1} N_s^{(j)}
\end{equation}
In the illustration of Figure \ref{fig_ros2}, $W=2$ for the top path and $W=18$ for the bottom path. $W$ in fact corresponds to the number of branches that would have been present, had we been using the BG scheme. As well as the generation of paths, we shall also discuss the computation of the probabilities $P(\lambda_{i+1}| \lambda_i)$. For this, it is important to note that the weighting factor for an {\em{incomplete}} transition path - {\em{i.e.}} one that connects $A$ to $\lambda_i$, is
\begin{equation}\label{wt2}
W_i = \prod_{j=0}^{i-1} N_s^{(j)}
\end{equation}

We now describe a practical scheme for sampling correctly weighted transition paths and for calculating the probabilities $P(\lambda_{i+1}| \lambda_i)$. The scheme (which we denote RB/M) uses  a Metropolis acceptance/rejection reweighting procedure \cite{daan}. An alternative ``Waste-Recycling'' approach (denoted RB/WR) is discussed in Appendix \ref{was}. The algorithm is as follows:\\

\noindent(1) Define values $W_t^{({\rm{o}})}$ and $W_t^{({\rm{n}})}$. For each interface $0 \le i < n$, define values $m_i=0$, $W_i^{({\rm{o}})}$, $W_i^{({\rm{n}})}$, $p_i^{({\rm{o}})}$, $p_i^{({\rm{n}})}$, $p_i^{cum}=0$. Define arrays of system configurations ${\mathcal{P}}^{({\rm{n}})}$ and ${\mathcal{P}}^{({\rm{o}})}$ in which to store transition paths.  \\
(2) Begin or continue a simulation run in the $A$ basin. When the system leaves $A$ and crosses $\lambda_0$, suspend this run. Denote the system configuration as $x_0$. Set  $i=0$.\\
(3) Increment $m_0 \to m_0+1$. Initiate $k_0$ trial runs from $x_0$. Continue each trial run until either $\lambda_1$ or $\lambda_A$ is reached. Calculate the number $N_s^{(0)}$ of trials which reach $\lambda_1$. Increment $p_0^{cum} \to p_0^{cum} + N_s^{(0)}/k_0$. Set $W_t^{({\rm{n}})} = N_s^{(0)}$ and $W_1^{({\rm{n}})}=N_s^{(0)}$.\\
(4) If $N_s^{(0)}>0$, choose one successful trial at random. Denote the final point of this trial as $x_1$. Add the configurations corresponding to this trial run into the array ${\mathcal{P}}^{({\rm{n}})}$. Set $i=1$. Otherwise (if $N_s^{(0)}=0$), return to step (2).\\ 
(5) Increment $m_i \to m_i+1$. Initiate $k_i$ trial runs from $x_i$. Continue each trial run until either $\lambda_{i+1}$ or $\lambda_A$ is reached. Calculate the number $N_s^{(i)}$ of trials which reach $\lambda_{i+1}$. Set $p_i^{({\rm{n}})}=N_s^{(i)}/k_i$. If $N_s^{(i)}>0$, select one successful trial at random and denote the final point of this trial as $x_{i+1}$.\\
(6) If $m_i=1$, set $p_i^{({\rm{o}})}=p_i^{({\rm{n}})}$ and $W_i^{({\rm{o}})}=W_i^{({\rm{n}})}$. If $m_i>1$, draw a random number $0 < r < 1$. If $r < W_i^{({\rm{n}})}/W_i^{({\rm{o}})}$, set $p_i^{({\rm{o}})}=p_i^{({\rm{n}})}$ and $W_i^{({\rm{o}})}=W_i^{({\rm{n}})}$. If $r > W_i^{({\rm{n}})}/W_i^{({\rm{o}})}$, $p_i^{({\rm{o}})}$ and $W_i^{({\rm{o}})}$ remain unchanged. \\
(7) Increment $p_i^{cum} \to p_i^{cum} + p_i^{({\rm{o}})}$. Increment $W_t^{({\rm{n}})} \to W_t^{({\rm{n}})}*N_s^{(i)}$. Set $W_{i+1}^{({\rm{n}})}=W_{i}^{({\rm{n}})}*N_s^{(i)}$.\\
(8) If $N_s^{(i)}=0$, return to step (2). Otherwise, increment $i \to i+1$ and repeat steps (5)-(8) until $i=n$.\\
(9) If $i=n$: if $m_n=1$, set $W_t^{({\rm{o}})}=W_t^{({\rm{n}})}$ and ${\mathcal{P}}^{({\rm{o}})}={\mathcal{P}}^{({\rm{n}})}$. Otherwise (if $m_n>0$), draw a random number $0 < r < 1$. If $r < W_t^{({\rm{n}})}/W_t^{({\rm{o}})}$, set $W_t^{({\rm{o}})}=W_t^{({\rm{n}})}$ and ${\mathcal{P}}^{({\rm{o}})}={\mathcal{P}}^{({\rm{n}})}$. If $r > W_t^{({\rm{n}})}/W_t^{({\rm{o}})}$, $W_t^{({\rm{o}})}$ and ${\mathcal{P}}^{({\rm{o}})}$ remain unchanged.\\
(10) The path ${\mathcal{P}}^{({\rm{o}})}$ is a member of the TPE and should be included in any analysis of the transition mechanism.\\
(11) Repeat steps (2)-(10) many times. \\
(12) For each interface $0 \le i < n$, calculate $P(\lambda_{i+1}|\lambda_i)=p_i^{cum}/m_i$.  The flux ${\overline{\Phi}}_{A,0}$ is given by $m_0/T$ where $T$ is the total length of the simulation run in the $A$ basin.\\

In this scheme, transition paths are generated by shooting $k_i$ trials from each interface $i$ and selecting one successful trial at random. $m_i$ denotes the number of paths to interface $i$ that have been generated.  When a complete path from $A$ to $B$ has been generated, its Rosenbluth weight $W_t^{({\rm{n}})}$, given by Eq.(\ref{wt1}), is compared to the Rosenbluth weight $W_t^{({\rm{o}})}$ of the previous complete path to be accepted (step (9)). The newly generated path is accepted if $W_t^{({\rm{n}})}/W_t^{({\rm{o}})}> r$ where $r$ is a random number between $0$ and $1$ (unless it is the first path to be generated ($m_n=1$), in which case it is always accepted). This Metropolis scheme has the effect of reweighting transition paths according to their Rosenbluth factors. The scheme also incorporates Metropolis acceptance/rejection steps at every interface, in step (6). This is necessary for correct calculation of the probabilities  $P(\lambda_{i+1}|\lambda_i)$, since the probability estimate $p_i=N_s^{(i)}/k_i$ obtained by firing $k_i$ trial runs from interface $\lambda_i$ must be reweighted by a factor $W_i$ given by Eq.(\ref{wt2}) which depends on the partial path leading from $A$ to $\lambda_i$. We note that the generation of the transition path continues to interface $i+1$ regardless of the outcome of the acceptance/rejection step at interface $i$. We also note that the ``previous partial path to be accepted'' at interface $i$ need not have any segments in common with the ``previous partial path to be accepted'' at any other interfaces.

In Appendix \ref{m2_app}, we demonstrate that this scheme indeed samples the TPE correctly, and we discuss the differences between this approach and the Rosenbluth method for the sampling of polymer configurations.

\subsection{Pruning}\label{sec_prune}
The analogy with polymer simulations also suggests a possible improvement to the efficiency of all three methods. By ``pruning'' trial paths which go in the direction of region $A$, we can avoid the computational expense of integrating  ``failed'' trials from interface $i$ all the way back to $\lambda_A$. In analogy with the Pruning method used for polymers  \cite{grassberger,daan},   trial paths which reach the previous interface at $\lambda_{i-1}$ from $\lambda_i$ are terminated with a certain probability. Surviving trials must be re-weighted to preserve the correct weights in the final TPE. The implementation of pruning in the context of these path sampling methods is described in Appendix \ref{prune}: for the genetic switch and polymer translocation problems described here, we find that although the results for $k_{AB}$ continue to be correct when pruning is used, no significant improvement in efficiency is achieved.

\section{Applications}

\subsection{Genetic switch}\label{gensw}

We now move on to a demonstration of the methods of Sections \ref{sec_ffs}, \ref{sec_bg} and \ref{ros_samp} for several non-equilibrium rare event problems. Our first test system is a set of chemical reactions comprising a symmetric bistable genetic switch, which is simulated using a kinetic Monte Carlo scheme. This is a non-equilibrium system whose dynamics does not obey detailed balance \cite{warren1,warren2,allen}. Examples of real genetic switches include the lysis-lysogeny switch of phage $\lambda$ \cite{ptashne} and the lac system of {\em{E. coli}} \cite{ozbudak}, as well as artificially engineered bacterial genetic switches \cite{gardner,isaacs}.

\begin{figure}[h]
\begin{tabular}{cccc|cc}
\multicolumn{3}{c}{} &\,\, & \multicolumn{1}{c}{\,$k_{f}$}&\multicolumn{1}{c}{$k_{b}$}\\
$2{\mathrm{A}} \rightleftharpoons {\mathrm{A}}_2$ &\,\,& $2{\mathrm{B}} \rightleftharpoons {\mathrm{B}}_2$ &\,\,&\, $5k$ & $5k$\\ 
${\mathrm{O}} + {\mathrm{A}}_2 \rightleftharpoons {\mathrm{O}}{\mathrm{A}}_2$ &\,\,& $ {\mathrm{O}} + {\mathrm{B}}_2 \rightleftharpoons {\mathrm{O}}{\mathrm{B}}_2$ &\,\,&\, $5k$ & $k$\\
$ {\mathrm{O}} \to {\mathrm{O}} + {\mathrm{A}}$ &\,\,& $ {\mathrm{O}} \to {\mathrm{O}} + {\mathrm{B}}$ &\,\,& \,$k$ & -\\
$ {\mathrm{O}}{\mathrm{A}}_2 \to {\mathrm{O}}{\mathrm{A}}_2 + {\mathrm{A}}$ &\,\,& ${\mathrm{O}}{\mathrm{B}}_2 \to {\mathrm{O}}{\mathrm{B}}_2 + {\mathrm{B}}$ & \,\,&\,$k$ & -\\
$ {\mathrm{A}} \to \emptyset$ &\,\,& ${\mathrm{B}} \to \emptyset$ &\,\,&\, $0.25k$ & -\\\\
\end{tabular}
\newline
{\vspace*{0.5cm}{\includegraphics[scale=0.4,clip=true]{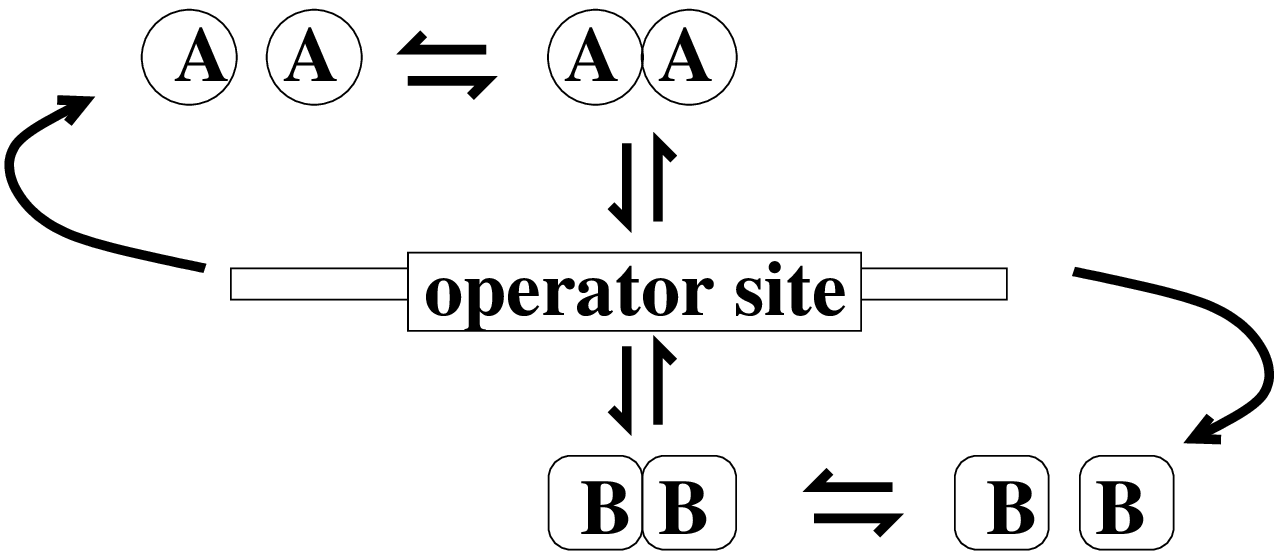}}}
\vspace*{-0.75cm}
\caption{Reaction scheme for the genetic switch. Proteins {\rm{A}} and  {\rm{B}} can dimerize and bind to the DNA at the operator site, {\rm{O}}. When ${\mathrm{A}}_2$ is bound to ${\mathrm{O}}$, ${\mathrm{B}}$ is not produced, and when ${\mathrm{B}}_2$ is bound to ${\mathrm{O}}$, ${\mathrm{A}}$ is not produced. Both proteins are degraded in the monomer form.  Forward and backward rate constants $k_{f}$ and $k_{b}$ are identical for A and B.\label{rs}}
\end{figure}

   Our model genetic switch is shown in Fig. \ref{rs}. The switch consists of a piece of DNA containing two genes ${\mathcal{A}}$ and ${\mathcal{B}}$, as well as a controlling operator site, ${\mathrm{O}}$. The two genes encode proteins $\rm{A}$ and $\rm{B}$, and we assume that (when  ${\mathrm{O}}$ is unoccupied) each of these proteins is produced from the DNA in a one-step process with rate constant $k$. In nature, of course, protein production is a complex multi-step process, the details of which we ignore.  Both proteins can dimerize and their dimers  $\rm{A}_2$ and $\rm{B}_2$ can bind to the operator site - however, only one dimer can be bound at any time. The binding of dimers to the operator site has the effect of controlling protein production - when $\rm{A}_2$ is bound to $O$, $\rm{A}$ is produced at rate $k$, but $\rm{B}$ is not produced. Likewise, when $\rm{B}_2$ is bound to $O$, $\rm{B}$ is produced at rate $k$, but $\rm{A}$ is not produced. Each dimer therefore blocks the production of the other protein. Both proteins are also removed (by enzymatic degradation or dilution due to cell growth) at a constant rate of $0.25k$. The genetic switch is bistable, having two steady states, one with a large number of $\rm{A}$ molecules, and few $\rm{B}$, and the other with a large number of $\rm{B}$ and few $\rm{A}$ molecules. Switching between these states (``flipping'') occurs due to stochastic fluctuations; the factors affecting the flipping rate have  been extensively investigated \cite{cherry,kepler,warren1,warren2}.

 We simulate the switch using the Gillespie algorithm \cite{gillespie}. This is  a widely-used kinetic Monte Carlo scheme for propagating chemical reactions. In each simulation step, a random number is drawn from the correct (exponential) distribution and used to choose the time at which the next reaction will occur, and another random number is used to determine which reaction this will be. The simulation time and numbers of molecules of all species are then updated accordingly. The phase space in these simulations is the number of molecules of each chemical species present in the system.  A full description of the simulation algorithm is given in Ref. \cite{gillespie}. An initial version of the results for the FFS method, as well as a discussion of  some interesting features of the TPE, was given in a previous publication \cite{allen}.

\begin{figure}[h]
\begin{center}
{\rotatebox{0}{{\includegraphics[scale=0.27,clip=true]{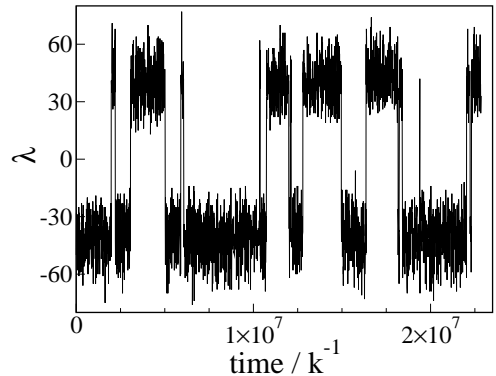}}}}\hspace{1cm}
\caption{$\lambda$ as a function of time (in units of $k^{-1}$) for a
typical simulation run.\label{gil1} }
\end{center}
\end{figure}

For the parameter $\lambda$, we choose $\lambda=N_{\rm{B}}-N_{\rm{A}}$, where $N_{\rm{A}}$ is the total number of ${\rm{A}}$ proteins, and $N_{\rm{B}}$ the total number of ${\rm{B}}$ proteins:
\begin{eqnarray}
&&N_{\rm{A}} = n_{\rm{A}} + 2\left(n_{\rm{A_2}}+n_{\rm{OA_2}}\right)\\\nonumber &&N_{\rm{B}} = n_{\rm{B}} + 2\left(n_{\rm{B_2}}+n_{\rm{OB_2}}\right)
\end{eqnarray}
$n_X$ being the number of molecules of species $X$. Figure \ref{gil1} shows $\lambda$ as a function of time for a typical brute-force simulation run (note that the unit of time in these simulations is $k^{-1}$). It is clear that  the system is indeed bistable, and that transitions are rapid in comparison to the waiting time between events. The parameters of Fig.\ref{rs} have been chosen to give a rather fast flipping rate, which can be measured using brute force simulation.  We define regions $A$ and $B$ by $\lambda_A=\lambda_0=-24$ and $\lambda_B=24$. A ``flip'' is considered to have occurred when the system enters region $B$, having come from $A$ ({\em{i.e.}} having $h_{\mathcal{A}}=1$, meaning it was in $A$ more recently than in $B$). To calculate $k_{AB}$, the integral $F(t) = \int_0^{t} dt'\, p(t')$ of the distribution $p(t)$ of times between flips was fitted to the Poisson function $F(t) = 1-\exp{[-k_{AB}t]}$, leading to a result  $k_{AB}=(9.4 \pm 0.2) \times 10^{-7} k$. This calculation was done over a total brute force simulation time of $9 \times 10^9 k^{-1}$, during which  8808 flips occurred.

\begin{table}[h]
\begin{center}
\begin{tabular}{cccccc}
& $f / k \times 10^{-2}$ &$P_B \times 10^{-5}$ &\,\, $k_{AB} / k \times 10^{-7}$ & $N_{st} \times 10^{11}$\\
BF & - & - & $\bf 9.4 \pm 0.2$&$14.8$\\
FFS & $1.221 \pm 0.005$ & $7.8 \pm 0.1$ & $9.4 \pm 0.2$&$1.1$\\
BG  & $1.212 \pm 0.006$ & $7.6 \pm 0.2$ & $9.3 \pm 0.2$&$0.5$\\
RB/M &  $1.220 \pm 0.004 $& $7.8 \pm 0.1 $& $9.4 \pm 0.1$&$1.8$\\
RB/WR & $1.223 \pm 0.004$& $7.7 \pm 0.2$ & $9.4 \pm 0.3$&$1.0$\\
\end{tabular}
\end{center}
\caption{Path sampling and brute force results for
$f={\overline{\Phi}}_{A,0}/{\overline{h}}_{\mathcal{A}}$,
$P(\lambda_n|\lambda_0)$ and $k_{AB}$. The brute force result is obtained
by fitting $F(t)$ as described in the text. $N_{st}$ is the approximate number of simulation steps performed in obtaining the result in the table.\label{gil_res}}
\end{table}

The results of calculations of the flipping rate using FFS, BG, RB with Metropolis acceptance/rejection (RB/M) and RB with Waste Recycling (RB/WR), for the same parameter set, are shown in Table \ref{gil_res}. In all cases, $\lambda_A=\lambda_0=-24$ and  $\lambda_B=\lambda_n=24$. We used $n=7$ and the interfaces were placed at $\lambda_i = \{-24,-22,-18,-15,-12,-9,-4,24\}$ ($0 \le i \le n$). The number of trials $k$ for the RB and BG methods was $k_i = \{6,5,4,4,5,5,4\}$ ($0 \le i < n$). For FFS, we chose $N_0=1000$ and the number $M$ of trials at each interface was $M_i = \{6000,5000,4000,4000,5000,5000,4000\}$ ($0 \le i < n$). The calculations were carried out as a series of  ``blocks'', each consisting of 1000 starting points for the RB and BG methods, and of one FFS run for FFS. Results were averaged over all blocks. Table \ref{gil_res} shows excellent agreement  with the brute force result for all the path sampling methods. The values of $P(\lambda_{i+1}|\lambda_i)$ were found to be: $P(\lambda_{i+1}|\lambda_i) = \{0.25,0.20,0.30,0.26,0.24,0.24,0.34\}$ for $0 \le i < n$. The approximate number of simulation steps performed in obtaining the result in Table \ref{gil_res} is also given: it is clear that all the path sampling methods are much more efficient than brute force simulation, even for this relatively rapidly flipping switch. In a previous publication \cite{allen}, it was demonstrated that the improvement in efficiency of FFS over brute force simulation was dramatically increased when the switch flipping rate was decreased. In this work, we have not attempted to optimize the number and positioning of interfaces, or the number of trials carried out at each interface. Table \ref{gil_res} does not, therefore, provide a reliable guide to the relative efficiency of the various path sampling methods. However, we can make the general observation that the compuational efficiency is of the same order of magnitude for all of the methods. This issue will be discussed in detail in a future publication \cite{long2}.

\subsection{Driven Polymer Translocation through a pore}\label{poltr}
Our second test system represents a simplified approach to the important problem of polymer transport through a nanopore. This is a widely occurring phenomenon: biological examples include protein translocation through pores, RNA transport across the nuclear membrane, and injection of genetic material by viruses, while technological applications include gene therapy and sequencing of DNA. In general, translocation does not occur in equilibrium, but in response to a driving force, such an electrical field or the action of motor proteins. An important difference to the genetic switch flipping of Section \ref{gensw} is that this is not a bistable system but rather an escape problem: translocation events occur only in one direction and after each event the system is re-equilibrated in its original configuration. 

\begin{figure}[h]
\begin{center}
{\rotatebox{0}{{\includegraphics[scale=0.65,clip=true]{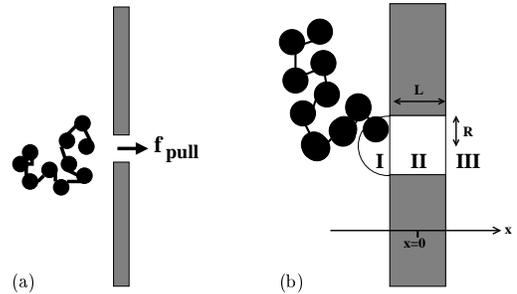}}}}
\caption{(a) An illustration of the polymer simulation. (b) A ``zoomed in'' view, showing the three regions used to define $\lambda$, in Eq. (\ref{lam_pol}).\label{polfig}}
\end{center}
\end{figure}

We have applied the path sampling methods of Section \ref{sec_ffs}, \ref{sec_bg} and \ref{ros_samp} to Langevin Dynamics simulations of the non-equilibrium, unidirectional translocation of a polymer through a pore.  The simulation setup is shown schematically in Figure \ref{polfig}. Our model polymer consists of $N$ monomers, each of which interacts with all other monomers {\em{via}} a spherical Lennard-Jones potential with parameters $\epsilon$ and $\sigma$:
\begin{equation}\label{lj_eq}
v_{lj}(r) = 4\epsilon \left[\left(\frac{\sigma}{r}\right)^{12}-\left(\frac{\sigma}{r}\right)^{6}\right]
\end{equation}
where $r$ is the distance between the monomers. This interaction is truncated at $r= 3\sigma$. Each monomer also interacts with its neighbours along the chain {\em{via}} a  linear spring potential (spring constant $s_s$) of the form:
\begin{equation}
v_{s}(r) = s_s(r-r_0)^2
\end{equation}
Here, $r$ is the distance between two neighbouring monomers and $r_0$ is the bond length.  The pore, of  radius $R$, embedded in a slab of width  $L$, is modelled by a repulsive Lennard-Jones potential with parameters $\epsilon_w$ and $\sigma_w$:
\begin{equation}\label{w_eq}
v_{w}(r) = 4\epsilon_w \left(\frac{\sigma_w}{r}\right)^{12}
\end{equation}
where $r$ is now the shortest distance between a monomer and the wall of the pore or slab. This interaction is also truncated at $r=3\sigma$.  

We do not aim at present to undertake a detailed study of the mechanism of polymer translocation. We therefore neglect the process by which the polymer arrives at the pore mouth, constraining the first monomer in the chain not to move far from the pore entrance on the left-hand side. This is achieved by applying a harmonic restraining force (spring constant $s_{hr}$) to the first monomer, of the form:
\begin{eqnarray}\label{harm}
f_{hr}(r_1) &= -s_{hr}(r_1-R) & r_1>R\\\nonumber
 &=0&{\rm{otherwise}}
\end{eqnarray}
In Eq.(\ref{harm}), $r_1$ is the distance of the first monomer from the point $(-L/2,0,0)$. The force acts along the vector connecting the first monomer to this point. If the first monomer is within a hemisphere of radius $R$ around the pore mouth, or is inside the pore,  or beyond the pore on the right-hand side, the restraining force is zero. 

To model the pulling of the polymer through the pore, we suppose that there exists some mechanism which exerts force on any monomers which are inside the pore. This force is, however, intermittent in time: the pore flips between states ON and OFF at rates $k_1$ (for the off $\to$ on transition) and $k_{-1}$ (for the on $\to$ off transition). When the pore is in the ON state, all monomers inside the pore experience a force $f_{pull}$ in the positive x direction. When the pore is in the OFF state, no pulling force is exerted. Although this model is not meant to represent any particular system, intermittent pulling forces of this kind might be produced by  motor proteins localized inside pores. The intermittent pulling force makes this a non-equilibrium system.

The monomers also experience stochastic forces due to the effects of solvent, and their dynamics is simulated according to the usual Langevin Dynamics algorithm:
\begin{equation}\label{bd1}
\dot{r}_{i\alpha}(t) = \frac{{p}_{i\alpha}(t)}{m}
\end{equation}
and
\begin{equation}\label{bd2}
\dot{p}_{i\alpha}(t) = -\xi {p}_{i\alpha}(t) + f_{i\alpha}(t) + \mathring{p}_{i\alpha}(t) 
\end{equation}
where ${r}_{i\alpha}$ is the i-th component of the position vector of monomer $i$, ${p}_{i\alpha}$ is the i-th component of its momentum vector and ${f}_{i\alpha}$ is the i-th component of the force acting on it due to the other monomers, the interactions with the wall, the pulling force and (for the first monomer only) the constraint force. The parameter $m$ is the monomer mass, $\xi$ is the friction constant, related to the diffusion constant $D$ by $\xi = k_B/mD$, and $\mathring{p}_{i\alpha}$ is a ``random force'' representing collisions with the solvent molecules and satisfying
\begin{equation}
\langle \mathring{p}_{i\alpha}(t)\mathring{p}_{i\beta}(0)\rangle = 2mk_BT\xi \delta(t) \delta_{\alpha \beta} 
\end{equation}
Equations (\ref{bd1}) and (\ref{bd2}) are integrated with a finite timestep $dt$ using the predictor-corrector-type algorithm given in the book of Allen and Tildesley \cite{allentildesley}. If, at a particular step, the state of the pulling force is OFF, it is changed to ON with probability $k_1 dt$, and if it is ON, it is changed to OFF with probability $k_{-1} dt$.  Once the entire polymer has passed through the pore ({\em{i.e.}} all monomers are located at $x>L/2$), we replace it in its   starting position (a pre-equilibrated configuration), re-equilibrate for $T_{eq}=100 \sigma^2/D$ and continue the simulation.

 The parameters of the simulation were chosen such that the monomers attract each other strongly and the polymer adopts a globular configuration before entering the pore. To enter the pore, the polymer is forced to adopt an energetically unfavourable extended configuration. This scenario could model protein translocation. From the point of view of our calculations, it has the advantage of ensuring that the waiting time between translocation events is long compared to the length of the events themselves.  The parameter values were: $N=10$, $R=2\sigma$, $L=2\sigma$, $dt=0.02 \sigma^2/D$,$r_0=0.5\sigma$,$s_s=2k_BT/\sigma^2$,$\epsilon=2.5k_BT$,$\epsilon_w=0.3k_BT$,$\sigma_w=1 \sigma$ $f_{hr}=5k_BT/\sigma^2$,$f_{pull}=1.0k_BT/\sigma$,$k_1=10 \sigma^{-2}D$, $k_{-1}= 1\sigma^{-2}D$ and $L_x=L_y=L_z=200\sigma$. Our units of length are taken to be the Lennard-Jones interaction parameter $\sigma$; units of mass are $m$, units of energy are $k_BT$ and the diffusion constant defines the units of time, which are $\sigma^2/D$.  The simulation box is cuboidal, with dimensions $L_x$, $L_y$, $L_z$; periodic boundary conditions were used in all three dimensions.

We define the parameter $\lambda$ in a rather trivial way. We consider three regions, illustrated in Fig. \ref{polfig}b: the hemispherical region of radius $R$ around the left-hand pore mouth (region I), the region inside the pore (region II), and the region outside the pore on the right-hand side (region III).  Taking $n_I$, $n_{II}$ and $n_{III}$ to be the numbers of monomers in the three regions, we define
\begin{equation}\label{lam_pol}
\lambda = \frac{n_I/4 + n_{II}/2 + n_{III}}{N}
\end{equation}
During the translocation process, $\lambda$ increases from a value of $\lambda \le 1/(4N)$, to unity. We note that expression  (\ref{lam_pol})  is chosen merely for convenience, and is not expected to reflect the true reaction mechanism. A simpler definition might be the number of monomers which have already translocated ($\lambda=n_{III}$) - although this would also lead to the correct value of $k_{AB}$, in practice it gives rather few crossings of the first interface and is therefore less efficient.  

\begin{figure}[h]
\begin{center}
{\rotatebox{0}{{\includegraphics[scale=0.27,clip=true]{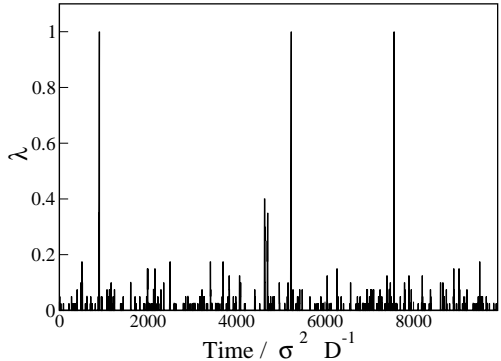}}}}
\caption{(a): $\lambda$ as a function of time (in units of $\sigma^2 D{-1}$) for a
typical brute force simulation run. \label{pol1} }
\end{center}
\end{figure}

Figure \ref{pol1} shows $\lambda$ as a function of time for a typical brute force simulation run. Translocation events occur rapidly compared to the waiting time between events. Defining an event to have occurred at the moment that the system crosses the interface $\lambda_n=1$, the waiting time distribution $p(t)$ can be measured and its integral $F(t) = \int_0^{t} dt'\, p(t')$ fitted to the Poisson function $F(t) = 1-\exp{[-k_{AB}t]}$, in order to measure $k_{AB}$. This resulted in a brute force measurement  $k_{AB}=(1.48 \pm 0.02) \times 10^{-4} D \sigma^{-2}$, obtained by simulating 5912 translocation events.

\begin{table}[h]
\begin{center}
\begin{tabular}{cccccc}
& $f \times 10^{-1}$ & $P_B \times 10^{-3}$ & $k_{AB} \times 10^{-4}$& $N_{st}\times 10^8$\\
BF & - & - & $\bf 1.48 \pm 0.02$ & $20.0$\\
FFS & $ 1.084\pm 0.006 $ & $ 1.36\pm 0.02 $ & $1.48\pm 0.02$  &$5.5$\\
BG  & $ 1.084\pm 0.006$ & $ 1.35 \pm 0.02$ & $ 1.47 \pm 0.02 $ &$3.2$\\
RB/M  & $ 1.086 \pm 0.006$& $ 1.31 \pm 0.03$& $ 1.43\pm 0.03 $ &$4.9$\\
RB/WR & $ 1.092 \pm 0.006$& $1.35\pm 0.03$ & $ 1.47\pm 0.03$ &$4.9$\\
\end{tabular}
\end{center}
\caption{Path sampling and brute force results for
$f={\overline{\Phi}}_{A,0}/{\overline{h}}_{\mathcal{A}}$,
$P(\lambda_n|\lambda_0)$ and $k_{AB}$.  Units of $f$ and $k_{AB}$ are $D\sigma^{-2}$. The brute force result is obtained
by fitting $F(t)$ as described in the text. Errors represent the standard error in the mean of a series of independent estimates.  $N_{st}$ is the approximate number of simulation steps performed in arriving at the result given in the table. \label{pol_res}}
\end{table}

The FFS, BG and RB methods (using both Metropolis acceptance/rejection and Waste Recycling) were applied to the polymer translocation problem, using the definition (\ref{lam_pol}) of $\lambda$. States $A$ and $B$ were defined by $\lambda_A=\lambda_0=0.025$ and $\lambda_B=\lambda_n=1.0$. We used $n=7$, with interfaces positioned at $\lambda_i=\{0.025,0.05,0.1,0.15,0.2,0.3,0.5,1.0\}$ for $0 \le i \le n$. For the FFS method, we used $N_0=500$ and the number of trials $M$ at each interface was  $M_i=\{1500,1500,1000,1000,1500,1500,550\}$ for $0 \le i < n$. For the BG method, the number of trials per point was $k_i=\{3,3,2,2,3,3,1\}$, while for both RB schemes it was $k_i=\{6,6,4,4,6,6,2\}$ ($0 \le i < n$). In all cases, averages were taken over a series of ``blocks'', each consisting of 500 starting points for the BG and RB methods, or of one FFS run. The results, given in table \ref{pol_res}, show  good agreement with the brute force results. The number $N_{st}$ of simulation steps used in the calculations is significantly lower for the path sampling techniques. 

\begin{table}[h]
\begin{center}
\begin{tabular}{cccccc}
& $f \times 10^{-2}$ & $P_B \times 10^{-5}$ & $k_{AB} \times 10^{-6}$& $N_{st}\times 10^8$\\
FFS & $ 9.88\pm 0.06 $ & $ 3.4\pm 0.1 $ & $3.4\pm 0.1$& $3.6$\\
BG  & $ 9.70\pm 0.06$ & $ 3.6 \pm 0.1$ & $ 3.5 \pm 0.1 $& $7.1$\\
RB/M  & $ 9.77 \pm 0.03$& $ 3.8 \pm 0.3$& $ 3.7\pm 0.3 $& $20.9$ \\
RB/WR & $ 9.83  \pm 0.03$& $3.4\pm 0.2$ & $ 3.3\pm 0.2$ & $17.6$\\
\end{tabular}
\end{center}
\caption{Path sampling and brute force results for
$f={\overline{\Phi}}_{A,0}/{\overline{h}}_{\mathcal{A}}$,
$P(\lambda_n|\lambda_0)$ and $k_{AB}$, for the polymer translocation problem with the altered parameter set. Units of $f$ and $k_{AB}$ are $D\sigma^{-2}$.  Errors represent the standard error in the mean of a series of independent estimates. $N_{st}$ is the approximate number of simulation steps performed in arriving at the result given in the table. \label{pol_res_slow}}
\end{table}

The parameter set given above was designed so as to allow calculation of the translocation rate by brute force simulation, in order to test the path sampling methods. The methods can also be used, of course, for much rarer transitions where brute force simulation is not feasible. We have also carried out calculations of $k_{AB}$ for an altered parameter set, which is as above,  except that the monomer-monomer Lennard-Jones interaction parameter $\epsilon$ is increased to $\epsilon=5k_BT$, and the wall-monomer interaction parameter $\epsilon_w$ becomes $\epsilon_w=1k_BT$. This implies very strong attraction between the monomers and very strong repulsion between the monomers and the pore. The same interfaces were used. For FFS, the same parameters were used: $N_0=500$ and  $M_i=\{1500,1500,1000,1000,1500,1500,550\}$. For the BG method, the number of trials per point was $k_i=\{4,5,3,4,7,10,2\}$, while for both RB schemes it was $k_i=\{12,15,9,12,21,30,6\}$. These parameters were chosen for convenience, but no systematic attempt was made at optimization; thus the results should not be used to compare efficiencies of the various path sampling methods, although once again, we see that the efficiency of all the methods is within about the same order of magnitude, with the RB method being somewhat less efficient.  The results for this rarer translocation problem are given in Table \ref{pol_res_slow}. Since the rate constant is 44 times smaller in this case, we can suppose that to obtain a brute force result of comparable accuracy, approximately $44\times20\times 10^8\approx 9 \times 10^{10}$ simulation steps would be required.

\section{Discussion}\label{dis}

In this paper, we have described three methods for the calculation of rates and the sampling of transition paths, for rare events in equilibrium or non-equilibrium stochastic dynamical systems in stationary state. What is the origin of the increased efficiency of these methods over brute force simulations?  A general characteristic of rare events is that  the system makes many ``failed attempts'', in which a fluctuation drives the system in the direction of $B$, for each ``successful'' transition from state $A$ to state $B$. In a brute force simulation, one does not capitalize on these failed attempts, but simply waits for the rare successful transition. In the methods described here, once the system crosses a particular interface, this configuration is stored and trial runs are used to try to extend the path to subsequent interfaces. The interfaces thus allow us to capitalize on those fluctuations that drive the system in the direction of $B$, since the system advances from one interface to the next in a ratchet-like manner. This is the origin of the increase in efficiency over brute force simulation. Of course, situations may arise in which the majority of the fluctuations in the direction of $B$ in fact lead into ``blind alleys'', rather than generating transition paths. This problem could perhaps be overcome by  again exploiting the analogy with polymer simulations to develop a scheme based on the Recoil Growth method \cite{daan}.

The approaches described in this paper differ greatly from existing  path sampling methods for rare events. The most widely used method for generating the Transition Path Ensemble is Transition Path Sampling (TPS) \cite{tps}. TPS samples the TPE using a dynamic Markov Chain Monte Carlo algorithm. Here, a new path is generated by shooting off trajectories in the forwards and backwards directions from a point in the old
path, after slightly changing its momentum co-ordinate. The new path is then accepted or rejected, usually via a Metropolis
acceptance/rejection criterion (which requires  knowledge of  the phase space density of the new initial point, meaning that TPS cannot be used for non-equilibrium systems). The acceptance criterion is optimized by  tuning the maximum momentum
displacement.    However, even with deterministic
dynamics, the Lyapunov instability of the system is often so large
that when the smallest momentum displacements possible are used, the trial
trajectories still diverge in a few picoseconds from the old ones; for
stochastic dynamics, the situation is likely to be worse. TPS alleviates this
problem by mainly shooting off trajectories near the top of the
barrier; however, this drastically hampers the relaxation of the
transition paths and, as a result, TPS is inefficient for transitions
that take longer than a few picoseconds. The
Lyapunov instability also explains why TPS cannot conveniently be
adapted to simulate non-equilibrium systems by only shooting in the
forward direction, in the manner of the methods described here: shooting in the forward direction from early points
in the transition paths is very unlikely to succeed. The non-equilibrium scheme
of Crooks and Chandler~\cite{crooks} is also expected to suffer from trajectory
divergence for multi-dimensional systems.  The methods described in this paper  suffer much less from these 
problems associated with the Lyapunov instability. This is because trial runs which are fired from interface $\lambda_{i}$ are only required to reach $\lambda_{i+1}$ in order to make a contribution to the path ensemble. If the distance between interfaces were to be very large, the Lyapunov instability might lead to problems in reaching  $\lambda_{i+1}$, but in this case, the interfaces can simply be positioned more closely.  These
methods should therefore prove useful for studying diffusive rare
events.

The schemes presented here use the same formulation for the rate constant as the Transition Interface Sampling (TIS) method of Van Erp {\em{et al}} \cite{vanerp,vanerp2005}. In TIS, however, paths from $A$ to interface $\lambda_i$ are sampled using the ``shooting'' methodology of TPS. Although TIS is generally  more efficient than TPS for rate constant calculations, like TPS, it cannot be used for non-equilibrium systems, since knowledge of the phase space density is required. TIS also suffers from the Lyapunov instability in the same way as TPS and is therefore only suitable for very short transition paths.

Other schemes have also been proposed which use a series of interfaces between regions $A$ and $B$, including  Partial Path Transition Interface Sampling \cite{moroni,moroniphd} and Milestoning \cite{faradjian}. These methods assume  that the points in the TPE at the interfaces are distributed according to the stationary phase space density - for example, the Boltzmann distribution. This allows them to be used for diffusive problems where transition paths are long. However, this assumption is unjustified in many cases \cite{tenwolde}, even for equilibrium problems. Moreover, these methods cannot be used for non-equilibrium problems where the phase space density is unknown.

For the methods described in this paper, the use of interfaces does not involve any assumptions about the transition mechanism, or about the transition paths. The role of the interfaces is simply to improve the efficiency of the sampling; they have no effect on the transition paths that are obtained. This is because the final points of the trial runs from interface $i-1$ are used as the initial points of the trials from interface $i$, so that the correct dynamics of the system is preserved throughout the transition path. The points at the interfaces are not assumed to follow the steady state phase space distribution. In fact, for the genetic switch, we find that the distribution of points at the interfaces is very far from the steady state one \cite{allen}.

It is interesting to make some general points on the nature of the path sampling in the different schemes discussed here. The FFS and BG methods proposed here are examples of static
Monte Carlo schemes, in which new paths are generated independently of
previous paths. The RB/M scheme could be interpreted
as a dynamic Markov Chain Monte Carlo algorithm, since newly generated  paths are compared with previously generated ones. However,
new trajectories are here generated from scratch, in contrast to
most dynamic importance sampling algorithms (including TPS), where previous paths are used to generate new ones. The methods described here have the general advantage of static schemes that they
are less likely to get stuck in particular regions of state
space, a common problem in dynamic importance sampling
schemes. 

In this paper, we have demonstrated that all three of the methods provide a dramatic improvement in efficiency over brute force simulations, for calculations of the rate constant. For the problems studied here, the efficiency of all three methods was roughly of the same order of magnitude, with the RB method being slightly less efficient. A much more detailed study of the efficiency of the methods, in which analytical expressions are derived for the computational cost of the three algorithms and for the statistical accuracy of the resulting estimates of the rate constant, will be presented in a future publication \cite{long2}. This should allow systematic optimization of the choice of parameters, for particular methods applied to particular problems. The choice of which method to use may depend not only on the computational efficiency with which the rate constant can be calculated, but also on practical issues such as the fact that the BG and RB methods require less storage of system configurations than FFS. In cases where one is interested in analysing the TPE to obtain information on the transition mechanism, the RB method may be preferable, since it generates unbranched transition paths in a convenient, one-at-a-time fashion.

The methods described in this paper are only suitable for stochastic dynamical schemes, since they rely on the fact that many trials can be fired from one initial point. Many rare events are simulated using Molecular Dynamics, which is generally entirely deterministic. Is it possible to use schemes of this type for Molecular Dynamics simulations? Our view is that it is indeed possible, if a weak Andersen thermostat \cite{andersen1980} is used as a noise generator. This approach was used by Bolhuis to apply TPS to diffusive barrier crossings \cite{bolhuis2003}.  As long as the noise source does not increase the timescale of the Lyapunov divergence, it is unlikely to disturb the dynamics of the system. Further investigation in this direction is planned. 

Finally, the methods as formulated here are suitable for non-equilibrium systems in stationary state {\em{i.e.}} systems where detailed balance is not obeyed, there are fluxes in phase space and the phase space density is not known, but nevertheless the average properties of the system are time-independent. These conditions apply to a wide class of systems which have not previously been accessible to rare event simulations. However, another very interesting class of non-equilibrium rare events occurs in systems that are not in stationary state - for example, systems with a time-dependent external driving force. In future work, we aim to investigate under what circumstances these methods can be used for time-dependent problems.

\begin{acknowledgments}
The authors thank  Peter Bolhuis, Eduardo Sanz, Chantal Valeriani and Patrick B. Warren for many useful discussions. We also thank Chantal Valeriani for her very helpful reading of the manuscript and Ovidiu Radulescu for pointing out to us Ref. \cite{altamirano}.  This work is part of the research program of the ``Stichting voor Fundamenteel Onderzoek der Materie (FOM)'', which is financially supported by the ``Nederlandse Organisatie voor Wetenschappelijk Onderzoek (NWO)''. R.J.A. was funded by the European Union Marie Curie program. 
\end{acknowledgments}

\appendix
\section{Justification of the algorithms}\label{theor_j}

\subsection{Averaging of probabilities}\label{app_prob}
In this section, we comment on expressions (\ref{eq1}) and (\ref{eq2}) for the rate constant $k_{AB}$. Eq.(\ref{eq1}) states that $k_{AB}$ is the time-averaged flux ${\overline{\Phi}}_{A,n}$ of trajectories reaching $\lambda_n$, coming from $A$, per unit of time that the system spends in state $h_{\mathcal{A}}=1$. This is then equal to the time-averaged flux ${\overline{\Phi}}_{A,0}$ of trajectories crossing $\lambda_0$ for the first time since leaving $A$, multiplied by the probability $P(\lambda_n|\lambda_0)$ that any one of these trajectories will subsequently reach $\lambda_n=\lambda_B$, before returning to $A$. Eq.(\ref{eq2}) states that for a particular trajectory, $P(\lambda_n|\lambda_0)$ is equal to the probability of reaching $\lambda_1$ from $\lambda_0$, then, given that $\lambda_1$ has been reached, of subsequently reaching $\lambda_2$, and so on. In the Branched Growth method, $P(\lambda_n|\lambda_0)$ is indeed estimated for individual trajectories: for a particular starting point at $\lambda_0$, the product $\prod_{i=0}^{n-1}P(\lambda_{i+1}|\lambda_{i})$ is explicitly evaluated by creating a branching tree of paths and counting the number of branches that reach $\lambda_n$. An average is then taken of this estimate over many such branching paths. In the Branched Growth method, therefore, we obtain
\begin{equation}\label{ross2}
P(\lambda_n|\lambda_0)_{BG} = \langle \prod_{i=0}^{n-1}P(\lambda_{i+1}|\lambda_{i}) \rangle_{\lambda_0}
\end{equation}
where the notation $\langle  \rangle_{\lambda_0}$ denotes an average over all paths which begin from $\lambda_0$. 

In the FFS and Rosenbluth methods, however, averages are taken over the estimates of $P(\lambda_{i+1}|\lambda_{i})$ for each interface $i$, and these averages are multiplied: 
\begin{equation}\label{ross1}
P(\lambda_n|\lambda_0)_{FFS/RB} = \prod_{i=0}^{n-1} \langle P(\lambda_{i+1}|\lambda_{i}) \rangle_{\lambda_i} 
\end{equation}
In Eq. (\ref{ross1}),  $\langle  \rangle_{\lambda_i}$ denotes an average over all paths which extend from $A$ to $\lambda_i$. We now demonstrate that Eqs (\ref{ross2}) and (\ref{ross1}) are consistent. 

Beginning with Eq. (\ref{ross2}), we multiply and divide by $\langle \prod_{i=0}^{n-2}P(\lambda_{i+1}|\lambda_{i}) \rangle_{\lambda_0}$:
\begin{eqnarray}\label{ross3}
\nonumber P(\lambda_n|\lambda_0)_{BG} &=& \frac{\langle P(\lambda_{n}|\lambda_{n-1}) \times \prod_{i=0}^{n-2}P(\lambda_{i+1}|\lambda_{i})  \rangle_{\lambda_0}}{\langle \prod_{i=0}^{n-2}P(\lambda_{i+1}|\lambda_{i}) \rangle_{\lambda_0}}\\ && \times \langle \prod_{i=0}^{n-2}P(\lambda_{i+1}|\lambda_{i}) \rangle_{\lambda_0}\\\nonumber &=& \langle  P(\lambda_n|\lambda_{n-1}) \rangle_{\lambda_{n-1}}\times \langle \prod_{i=0}^{n-2}P(\lambda_{i+1}|\lambda_{i}) \rangle_{\lambda_0} 
\end{eqnarray}
since $\prod_{i=0}^{n-2}P(\lambda_{i+1}|\lambda_{i})$ is the weighting factor for a particular path that starts from $\lambda_0$, in the ensemble of paths that connect $\lambda_0$ to $\lambda_{n-1}$. We now multiply and divide by  $\langle \prod_{i=0}^{n-3}P(\lambda_{i+1}|\lambda_{i}) \rangle_{\lambda_0}$: 
\begin{eqnarray}\label{ross4}
\nonumber P(\lambda_n|\lambda_0)_{BG}  &=& \langle  P(\lambda_n|\lambda_{n-1}) \rangle_{\lambda_{n-1}} \times  \langle \prod_{i=0}^{n-3}P(\lambda_{i+1}|\lambda_{i}) \rangle_{\lambda_0}\\\nonumber & \times &\frac{\langle P(\lambda_{n-1}|\lambda_{n-2}) \prod_{i=0}^{n-3}P(\lambda_{i+1}|\lambda_{i}) \rangle_{\lambda_0}}{\langle \prod_{i=0}^{n-3}P(\lambda_{i+1}|\lambda_{i}) \rangle_{\lambda_0}}\\\nonumber &=& \langle  P(\lambda_n|\lambda_{n-1}) \rangle_{\lambda_{n-1}}\times \langle  P(\lambda_{n-1}|\lambda_{n-2}) \rangle_{\lambda_{n-2}}\\\nonumber &&\times \langle \prod_{i=0}^{n-3}P(\lambda_{i+1}|\lambda_{i}) \rangle_{\lambda_0}
\end{eqnarray}
Extending this analysis, we arrive at the result that $P(\lambda_n|\lambda_0)_{BG} = P(\lambda_n|\lambda_0)_{FFS/RB}$.

\subsection{Weights of paths}
In this section, we show that all three methods sample the true transition path ensemble (TPE): {\em{i.e.}} that transition paths are sampled with the correct weights. We define the TPE to be all paths that would be obtained in an infinitely long brute force simulation run, which obey the conditions that their first point is in $A$, their last point is in $B$, and all other points lie between $A$ and $B$. These paths can consist of any number $N$ of simulation steps. The weight of a particular transition path in the TPE is:
\begin{eqnarray}\label{app1}
\mathcal{P}(\{x\})= && C\theta(\lambda_A-\lambda(x_0))\rho (x_0) \\\nonumber && \times \prod_{i=0}^{N-2}p_{i,i+1} \theta(\lambda(x_{i+1})-\lambda_A)\theta (\lambda_{B}-\lambda(x_{i+1}))\\\nonumber && \times p_{N-1,N}\theta(\lambda(x_N)-\lambda_{B}) 
\end{eqnarray}
where $p_{i,i+1} = p(x_i\to x_{i+1})$, the probability of making a step from point $x_i$ to point $x_{i+1}$, $\rho_0(x_0)$ is the steady-state phase space density of the first point in the path, which is in region $A$, and C is a normalization constant. The first term in Eq.(\ref{app1}) is the phase space density of the initial point; the $\theta$-function ensures that point $x_0$ lies in $A$. The next term is a product over all simulation steps $i$ in the path, except the last point: $p_{i,i+1}$ is the probability of taking a particular simulation step and the $\theta$-functions ensure that point $x_{i+1}$ lies between $\lambda_A$ and $\lambda_B$. The final $\theta$-function ensures that the last point in the transition path lies in the $B$ region.
 
\begin{figure}[h]
\begin{center}
{\rotatebox{0}{{\includegraphics[scale=0.65,clip=true]{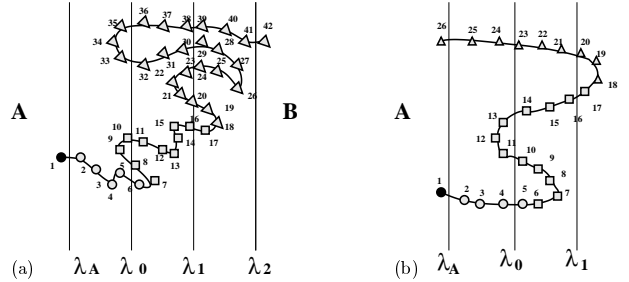}}}}
\caption{Illustration of the division of a path into partial paths. (a) A path which begins in $A$ and reaches $B$. Points 2-6 belong to the partial path $Y_{-1}$, points 7-17 to $Y_0$ and points 18-42 to $Y_2$. (b) A path which begins and ends in $A$. Partial paths are coded as follows: $Y_{-1}$ open circles, $Y_0$ squares, $Y_1$  triangles.\label{pp_fig} }
\end{center}
\end{figure}

We now divide the transition path into a series of partial paths. A partial path $Y_j$, consisting of a successive set of points $\{y^{(j)}_1 \dots y^{(j)}_{N_j}\}$, is defined to be a part of a trajectory (the whole trajectory being $\{x_0 \dots x_N\}$), which begins just after the trajectory crosses interface $\lambda_j$ for the first time, and ends just after it crosses either $\lambda_{j+1}$ or $\lambda_A$. The first partial path is denoted $Y_{-1}$. This begins just after the trajectory leaves $A$ and ends just after it crosses $\lambda_0$ for the first time, or returns to $A$.  Figure \ref{pp_fig} illustrate the division of two different trajectories into partial paths.  In Figure \ref{pp_fig}a, for example, $y^{(-1)}_1 = x_2$, $y^{(-1)}_{N_{-1}} = y^{(-1)}_5 = x_6$, $y^{(0)}_1 = x_7$, $y^{(0)}_{N_0} = y^{(0)}_{11} = x_{17}$,  $y^{(1)}_1 = x_{18}$, $y^{(1)}_{N_1} = y^{(1)}_{25} = x_{42}$.  We also define a ``success'' function, $\xi[Y_m]$, by
\begin{eqnarray}
\xi[Y_j]=1
\end{eqnarray}
if partial path $Y_j$ ends at $\lambda_{j+1}$ and 
\begin{eqnarray}
\xi[Y_j]=0
\end{eqnarray}
if  partial path $Y_j$ instead ends in region $A$. For example, in Figure  \ref{pp_fig}a, $\xi[Y_{-1}]=1$, $\xi[Y_0=1$ and $\xi[Y_1]=1$, whereas in Figure  \ref{pp_fig}b,  $\xi[Y_{-1}]=1$, $\xi[Y_0]=1$ and $\xi[Y_1]=0$.

\noindent Denoting the initial point of the path $y_A \equiv x_0$, we can now rewrite Eq.(\ref{app1}) as:
\begin{eqnarray}\label{app2}
\mathcal{P}(\{x\})= && C\theta(\lambda_A-\lambda(y_A))\rho_0(y_A)\\\nonumber && \times \prod_{j=-1}^{n-1} \bigg[ \prod_{i \in Y_j}p_{i,i+1} \bigg]\xi[Y_j]
\end{eqnarray}
where the innermost product is now over all points in the transition path which belong to partial path $Y_j$ - the $\theta$-functions of Eq.(\ref{app1}) are implicit in the definition of $Y_j$. The factor $\xi[Y_j]$ ensures that each partial path reaches the next interface. The final step $x_{N-1}\to x_N$ is included in partial path $Y_{n-1}$.

\subsection{The FFS and BG methods}\label{ffs_app}
The FFS and BG methods begin with a simulation run in the basin of attraction of $A$, from which points are collected immediately after the simulation crosses $\lambda_0$. The probability distribution for the partial paths that connect $A$ to $\lambda_0$ is:  
\begin{equation}\label{com1}
\mathcal{P}_{-1}(\{x\}) = C_{-1}\theta(\lambda_A-\lambda(y_A))\rho_0(y_A) \bigg[\prod_{i \in Y_{-1}}p_{i,i+1}\bigg]\xi[Y_{-1}]
\end{equation}
Here, the $\theta$-function ensures that the initial point $y_A$ lies in region $A$. $\rho_0(y_A)$ is the steady-state phase space density for point $y_A$. The product is over all the points in partial path  $Y_{-1}$, which connects region $A$ to $\lambda_0$, and the factor $\xi[Y_{-1}]$ ensures that the path reaches $\lambda_0$ rather than returning to $A$. Finally, $C_{-1}$ is a normalization constant. Having obtained the point at $\lambda_0$, trial runs are then used to extend the transition path to subsequent interfaces.  In FFS, points are selected at random from a collection at $\lambda_0$, while in BG, $k_0$ trials are run from each point at $\lambda_0$. However, this makes no difference to the  probability distribution for the resulting paths that connect $\lambda_A$ to interface j, which is: 
\begin{eqnarray}\label{com2}
\mathcal{P}_j(\{x\}) = && C_j\theta(\lambda_A-\lambda(y_A))\rho_0(y_A) \\\nonumber && \times \prod_{m=-1}^{j-1} \bigg[\prod_{i \in Y_m}p_{i,i+1}\bigg]\xi[Y_m]
\end{eqnarray}
Once again, the inner product is over all points that form part of partial path $Y_m$. Extending this analysis to the $n$-th interface, we obtain the result that the FFS and BG methods sample paths according to the correct distribution function, given by Eq.(\ref{app2}).

\subsection{The RB method}\label{m2_app}
We now turn to the Rosenbluth method, implemented with the Metropolis acceptance/rejection scheme (RB/M), as described in Section \ref{ros_samp}. We show that this method generates paths with the correct weights, as given by Eq.(\ref{app2}), and we point out some differences between the RB/M method and the Rosenbluth procedure usually used for polymer sampling \cite{daan}.

In the RB/M method, the Rosenbluth weights $W_t^{(\rm{n})}$ and $W_t^{(\rm{o})}$, which are compared in the acceptance/rejection step, depend on {\em{all}} the trial runs which were used to produce the paths. The acceptance/rejection procedure therefore depends on all trial runs, not just the ones that are selected and form part of the transition path. In order to demonstrate the validity of the method, we consider the probability of generating and accepting a particular ``decorated transition path'' - by which we mean a transition path from $A$ to $B$, {\em{together with its  $k_j-1$ attendant unselected trials for each interface j}}.

Let us suppose that we have reached interface $\lambda_j$ in the RB path generation procedure.   The probability of generating a particular trial run (or ``trial partial path'') $Y_j^b$ to $\lambda_{j+1}$ or $\lambda_A$ is:
\begin{equation}\label{app_ros1}
P^{gen}[Y_j^b] = D_j \prod_{i \in Y_j^b}p_{i,i+1}
\end{equation}
where $D_i$ is a normalization constant and the product is over all steps in the trial run $Y_j^b$. Having generated a set of $k_j$ trial runs, the probability of selecting a particular one, $Y_j^{*}$, to extend the chain to the next interface is:
\begin{equation}\label{app_ros2}
P^{sel}[Y_j^{*}] = \frac{\xi[Y_j^{*}]}{\sum_{b=1}^{k_j} \xi[Y_j^b]}
\end{equation}
where the index $b$ runs over all the $k_j$ generated trial runs $Y_j^b$. We now consider the generation of a new decorated transition path, consisting of a chain of partial paths $Y_j$ for $-1 \le j < n$. The probability of obtaining a particular path leading from $A$ to $\lambda_0$ is, as in Eq.(\ref{com1}):
\begin{equation}
\mathcal{P}_{-1}(\{x\}) = C_{-1}\theta(\lambda_A-\lambda(y_A))\rho_0(y_A) \bigg[\prod_{i \in Y_{-1}}p_{i,i+1}\bigg]\xi[Y_{-1}]
\end{equation}
We then shoot $k_j$ trial runs at each interface $\lambda_j$, for $0 \le j < n$. At each interface, we denote the trial run that is selected by an asterisk and the $k_j-1$ trials which are not selected  by the index $b'$. The probability of generating a particular decorated transition path, consisting of selected trial paths $Y_j^*$ and unselected trial runs $Y_j^{b'}$ is:
\begin{eqnarray}\label{app_ros4}
&& \nonumber{P}^{gen}({\rm{n}}) = C' \theta(\lambda_0-\lambda(y_A)) \rho_0(y_A) \bigg[\prod_{i \in Y_{-1}}p_{i,i+1}\bigg]\xi[Y_{-1}] \\ && \,\,\, \times \prod_{j=0}^{n-1}  \bigg[{P}^{gen}[Y_j^*] {P}^{sel}[Y_j^*]\prod_{b'=1}^{k_j-1}{P}^{gen}[Y_j^{b'}] \bigg] 
\end{eqnarray}
Having generated a decorated transition path (here denoted ${\rm{n}}$), we now compare its Rosenbluth factor to that of the last decorated transition path that was accepted (here denoted ${\rm{o}}$). The probability ${P}^{acc}({\rm{o\to n}})$ of acceptance obeys the relation:
\begin{equation}\label{app_ros5}
\frac{{P}^{acc}({\rm{o \to n}})}{{P}^{acc}({\rm{n \to o}})} = \frac{W_t^{({\rm{n}})}}{W_t^{({\rm{o}})}}
\end{equation}
where $W_t^{({\rm{n}})}$ and $W_t^{({\rm{o}})}$ are the Rosenbluth factors:
\begin{equation}
W_t^{({\rm{n}})} = \prod_{j=0}^{n-1} {\sum_{b=1}^{k_j} \xi[Y_j^b({\rm{n}})]} \qquad ; \qquad W_t^{({\rm{o}})} = \prod_{j=0}^{n-1} {\sum_{b=1}^{k_j} \xi[Y_j^b({\rm{o}})]}
\end{equation}
where the index $b$ runs over all (selected and unselected) trial runs at interface $j$ which belong to the decorated transition path. 

The flow of probability during the path sampling procedure from decorated path ${\rm{o}}$ to decorated path ${\rm{n}}$ is given by:
\begin{equation}\label{app_ros6}
{\mathcal{K}}({\rm{o \to n}}) = {\mathcal{N}}({\rm{o}}){{P}}^{gen}({\rm{n}}) {{P}^{acc}({\rm{o \to n}})}
\end{equation}
where ${\mathcal{N}}({\rm{o}})$ is the weight of the old augmented path in our ensemble. When our sampling reaches a steady state, detailed balance will be obeyed in the space of decorated paths: 
\begin{equation}\label{app_ros7}
{\mathcal{K}}({\rm{o \to n}}) = {\mathcal{K}}({\rm{n \to o}})
\end{equation}
Substituting Eqs (\ref{app_ros1})-(\ref{app_ros6}) into (\ref{app_ros7}), we find that:
\begin{eqnarray}
 && \frac{{\mathcal{N}}({\rm{n}})}{{\mathcal{N}}({\rm{o}})}= \frac{\theta(\lambda_A-\lambda(y_A({\rm{n}})))\rho_0(y_A({\rm{n}}))}{\theta(\lambda_A-\lambda(y_A({\rm{o}})))\rho_0(y_A({\rm{o}})) }\\\nonumber && \times \frac{\xi[Y_{-1}({\rm{n}})] \prod_{i \in Y_{-1}({\rm{n}})}p_{i,i+1}}{\xi[Y_{-1}({\rm{o}})]\prod_{i \in Y_{-1}({\rm{o}})}p_{i,i+1}}\\\nonumber&& \times \frac{  \prod_{j=0}^{n-1} {P}^{gen}[Y_j^*({\rm{n}})] \xi[Y_j^*({\rm{n}})]\prod_{b'=1}^{k_j-1} {P}^{gen}[Y_j^{b'}({\rm{n}})] }{ \prod_{j=0}^{n-1} {P}^{gen}[Y_j^*({\rm{o}})] \xi[Y_j^*({\rm{o}})]\prod_{b'=1}^{k_j-1} {P}^{gen}[Y_j^{b'}({\rm{o}})]}
\end{eqnarray}
from which we can conclude that a particular decorated path is sampled by the RB/M method with weight:
\begin{eqnarray}\label{app_ros8}
\nonumber  && {\mathcal{N}}(\{x\}) =  \theta(\lambda_A-\lambda(y_A))\rho_0(y_A)\xi[Y_{-1}] \prod_{i \in Y_{-1}}p_{i,i+1}\\&& \,\,\, \times  \prod_{j=0}^{n-1} {P}^{gen}[Y_j^*] \xi[Y_j^*]\prod_{b'=1}^{k_j-1} {P}^{gen}[Y_j^{b'}] 
\end{eqnarray}
where again, the index $b'$ denotes unselected trial runs. We would now like to know the weight with which a particular {\em{undecorated}} transition path is sampled in the $RB/M$ method. This weight is given by the sum of ${\mathcal{N}}(\{x\})$, taken over all decorated paths which have identical {\em{backbone chains}}: {\em{i.e.}} which represent identical transition paths, decorated by different sets of unselected trial runs. We know, however, that  
\begin{equation}
\sum^{''} \prod_{b'=1}^{k_i-1} {P}^{gen}[Y_i^{b'}] = 1
\end{equation}
where $\sum^{''}$ denotes a sum over all possible combinations of $k_i-1$ unselected trials from interface $i$. Taking this sum over the distribution function of Eq. (\ref{app_ros8}) and substituting in Eq.(\ref{app_ros1}), we find that the Rosenbluth method indeed samples transition paths  with the correct weight (\ref{app2}). 

The RB/M method described in this paper differs from the well-used Rosenbluth technique for polymer sampling \cite{daan}. There, the Rosenbluth factor of the newly generated polymer configuration is not compared to that of the previously accepted configuration, but rather to that of a randomly chosen chain from the system.   Moreover, the Rosenbluth factor of this chain must be re-calculated (by generating a new set of trial moves) when the chain is selected, rather than being stored when the configuration was first generated. This is necessary in the case of polymers because of the interactions between polymer chains, which depend on the current state of the system.  The RB/M technique of Section \ref{ros_samp}, which is much less computationally intensive, is appropriate for path sampling because of the absence of interactions between different transition paths.

\section{``Waste-Recycling''}\label{was}
In Section \ref{ros_samp}, we described the implementation of the Rosenbluth path sampling scheme, with a Metropolis-like acceptance/rejection procedure for reweighting the paths. Correct reweighting can also be achieved using an alternative approach, in which ensemble averages are computed over all generated paths, taking explicit account of their weights. This scheme, known as ``Waste-Recycling'', was originally proposed by Frenkel \cite{frenkel}, in a Monte-Carlo simulation context. 

Let us suppose that we wish to compute the average of a quantity $X$ for paths in the TPE. We know that paths from $A$ to $B$ which are generated by the Rosenbluth scheme should be weighted according to their Rosenbluth factors $W = \prod_{j=0}^{n-1} N_s^{(j)}$. We could, of course, simply compute  
\begin{equation}\label{wr1}
\langle X \rangle_{TPE} = \frac{\sum_{b=1}^{Q} W^{(b)} X^{(b)}}{\sum_{b=1}^{Q} W^{(b)}}
\end{equation}
where the index $b$ refers to the individual paths which are generated and $Q$ is the total number of generated paths.  The problem with this is that, as the path sampling proceeds, both  the numerator and denominator of Eq. (\ref{wr1}) will increase in proportion to the number of paths sampled. At the end of a long sampling run, one will be faced with the problem of dividing two enormously large numbers. The  ``Waste Recycling'' scheme avoids this problem. 

In order to use Waste Recycling to obtain the probabilities $P(\lambda_{i+1}|\lambda_i)$, as well as $\langle X \rangle_{TPE}$ for any chosen property $X$ of the transition paths, the following procedure is used:\\

\noindent(1)Choose a number $n_c$ (typically, $n_c \approx 3-10$). Define  values $X^{cum}=0$ for all properties $X$ of the TPE which one wishes to compute. For each interface ($0 \le i \le n$), define values $m_i=0$, $c_i=1$, $p_i^{cum}=0$ and arrays $W_i$ and $p_i$, of size $n_c$. Define an array of transition paths ${\mathcal{P}}$, also of size $n_c$.  \\
(2) Begin or continue a simulation run in the $A$ basin. When the system leaves $A$ and crosses $\lambda_0$, suspend this run. Denote the system configuration as $x_0$. Set $i=0$, $W_0[c_0]=1$.\\
(3) Initiate $k_i$ trial runs from $x_i$. Continue each trial run until either $\lambda_{i+1}$ or $\lambda_A$ is reached. Calculate the number $N_s^{(i)}$ of trials which reach $\lambda_{i+1}$. Set $p_i[c_i]=N_s^{(i)}/k_i$ and $W_{i+1}[c_{i+1}]=W_i[c_i]*N_s^{(i)}$. Increment $c_i \to c_i+1$.\\
(4) If $c_i<n_c+1$, continue to step (5). Otherwise ({\em{i.e.}} if $c_i=n_c+1$), increment $m_i \to m_i+1$ and 
\begin{equation}
p_i^{cum} \to p_i^{cum} + \frac{\sum_{b=1}^{n_c} p_i[b] W_i[b]}{\sum_{b=1}^{n_c} W_i[b]}
\end{equation}
Select one member of the array $W_i$ with probability $p_{sel}(b^*) = W_i[b^*]/\sum_{b=1}^{n_c}W_i[b]$. Set $W_i[1]=W_i[b^*]$, $p_i[1]=p_i[b^*]$ and $c_i=2$.\\
(5) If $N_s^{(i)}>0$, choose one successful trial at random. Denote the final point of this trial as $x_{i+1}$. Add the configurations corresponding to this trial run to the path ${\mathcal{P}}[c_t]$. Set $i \to i+1$. Otherwise (if $N_s^{(i)}=0$), return to step (2).\\ 
(6) Repeat steps (3)-(5) until $i=n$.\\
(7) If $i=n$: Increment $c_n \to c_n+1$.\\ 
(8) If $c_n<n_c+1$: return to step (2). Otherwise ({\em{i.e.}} if $c_n=n_c+1$): increment $m_n \to m_n+1$ and
\begin{equation}
X_{cum} \to X_{cum} + \frac{\sum_{b=1}^{n_c} W_n[b] X[{\mathcal{P}}[b]]}{\sum_{b=1}^{n_c} W_n[b]}
\end{equation}
Select one member of the array $W_n$ with probability $p_{sel}(b^*) = W_n[b^*]/\sum_{b=1}^{n_c}W_n[b]$. Set $W_n[1]=W_n[b^*]$, ${\mathcal{P}}[1]={\mathcal{P}}[b^*]$ and $c_n=2$.\\
(9) Repeat steps (2)-(8) many times. \\
(10) For each interface $0 \le i < n$, calculate $P(\lambda_{i+1}|\lambda_i)=p_i^{cum}/m_i$. Calculate  $\langle X \rangle_{TPE} = X_{cum}/m_n$. \\

The key idea of Waste Recycling is that one generates paths in ``groups'' - each group having $n_c$ members. Once a group of $n_c$ paths has been generated, the quantity $\left({\sum_{b=1}^{n_c} W_n[b] X[{\mathcal{P}}[b]]}\right)/\left({\sum_{b=1}^{n_c} W_n[b]}\right)$ is added to the cumulative average for the property $X$. One member of the group is then selected with probability proportional to its Rosenbluth weight $W$, to become the first member of the subsequent group. The algorithm described above above also includes separate ``grouping'' procedures for every interface: the index $c_i$ denotes the position of the partial path in the ``group'' connecting $A$ to $\lambda_i$ and $W_i[c_i]$ denotes the Rosenbluth factor of this partial path as given by Eq.(\ref{wt2}). Once a group of partial paths connecting $A$ to $\lambda_i$ contains $n_c$ members, the average $P(\lambda_{i+1}|\lambda_i)$ is incremented by $\left(\sum_{b=1}^{n_c} W_i[b] p_i[b]\right)/\left({\sum_{b=1}^{n_c} W_i[b]}\right)$ and one partial path is chosen with probability proportional to $W_i$ to be the first member of the next group.  This grouping procedure at each interface is necessary in order to correctly evaluate $P(\lambda_{i+1}|\lambda_i)$ using the partial path weights given by Eq.(\ref{wt2}).

In general, Waste Recycling can lead to very large increases in efficiency for Monte Carlo schemes in which a large set of possible moves (here paths) are generated, after which only one is accepted. This is not the case for our Rosenbluth path sampling scheme, where paths are generated one at a time. We therefore expect only a moderate, if any, increase in efficiency for the Waste Recycling scheme as compared to the Metropolis acceptance/rejection approach, for this particular application. In fact, as shown in Tables \ref{gil_res}, \ref{pol_res} and \ref{pol_res_slow}, the efficiency of the Waste Recycling and Metropolis acceptance/rejection schemes are comparable for the two test cases investigated here. Nevertheless, we have described and tested the scheme for the sake of clarity, completeness and future reference.

\section{Pruning}\label{prune}
For some problems, propagating trial paths from $\lambda_i$ back to $\lambda_A$ may be a major computational expense. In this case, computational efficiency could be enhanced  using  ``pruning'' - in analogy to the Pruned-Enriched Rosenbluth Method for polymer sampling \cite{grassberger,daan}.
 In the context of path sampling, this means that trial runs from $\lambda_i$ are not continued until they reach $\lambda_A$, but are rather terminated with probability $P_p$ on reaching the preceding interface $\lambda_{i-1}$. Surviving paths are re-weighted in order to maintain correct sampling of the TPE. We now discuss briefly the implementation of the pruning procedure for the three methods, and show for the polymer translocation problem of Section \ref{poltr} that the procedure leads to correct results for the rate constant.

\subsection{FFS}
The FFS algorithm proceeds as described in Section \ref{sec_ffs} until an interface $i$ is reached, such that $\lambda_{i-1}> \lambda_A$.  Each of the $N_i$ points in the collection at $\lambda_i$ is then assigned a weight  $f^{(i)}=1$. Selecting points at random, we carry out trial runs to $\lambda_{i+1}$. If a trial run arrives at $\lambda_{i-1}$, it is terminated and counted as a ``failure'' ({\em{i.e.}} it is counted as if it had reached $\lambda_A$), with probability $P_p^{(i-1)}$. The run continues with probability $1-P_p^{(i-1)}$, and its weight is multiplied by  $1/(1-P_p^{i-1})$. If it subsequently reaches $\lambda_{i-2}$, it is terminated  with probability $P_p^{(i-2)}$, and continues with probability $1-P_p^{(i-2)}$, with a weight which is now  $1/[(1-P_p^{i-1})(1-P_p^{i-2})]$. This process is continued until the trial run is terminated, it reaches  $\lambda_A$ or it finally arrives at $\lambda_{i+1}$.  The ``number of successes'', $N_s^{(i)}$ is now given by the sum of the weights of all  successful trials from $\lambda_i$ arriving at $\lambda_{i+1}$. On beginning the next trial run procedure, from $\lambda_{i+1}$ to $\lambda_{i+2}$, we  choose points from the collection at $\lambda_{i+1}$ with probability proportional to their weights $f^{(i)}$. Each of the new trial runs then begins with weight $f^{(i+1)}=1$. After performing $M_{i+1}$ trials, the  ``number of successes'' $N_s^{i+1}$ is the sum of the weights $f^{(i+1)}$ of all successful trials, and points at $\lambda_{i+2}$ are subsequently chosen according to their weights  $f^{(i+1)}$. Note that all trial runs begin with unit weight and not with the weight of their starting point in the collection at $\lambda_i$.

\subsection{Branched growth}
In the Branched Growth method, as described in Section \ref{sec_bg}, a branching ``tree'' of paths is created, in which $k_i$ trials are fired from each ``parent'' branch at interface $i$. Without pruning, the weight of each of the $k_i$ ``daughter'' branches is the weight of the ``parent'' branch, multiplied by $1/k_i$. When pruning is included, these weights are modified.  Suppose a trial run begins from interface $i$ with weight $h^{(i)}$. This weight $h^{(i)}$ will be equal to $1/\prod_{j=1}^{i-1} k_j$, multiplied by any factors due to pruning events that have occured during  the generation of the path from $\lambda_A$ to $\lambda_i$. Now suppose that this trial run does not proceed directly to $\lambda_{i+1}$, but rather goes back to  $\lambda_{i-1}$. It will then be terminated with probability $P_p^{(i-1)}$. However, let us suppose that it survives (with probability $1-P_p^{(i-1)}$). Its weight now becomes  $h^{(i)}/(1-P_p^{(i-1)})$. If it subsequently continues in the backward direction as far as $\lambda_{i-2}$ and survives the pruning procedure there, its weight will be $h^{(i)}/[(1-P_p^{(i-1)})(1-P_p^{(i-2)})]$, and so on. Due to the pruning procedure, not all branches reaching a particular interface will have the same weight [in the absence of pruning, the weight of all branches reaching $\lambda_i$ is $1/\prod_{j=1}^{i-1} k_j$]. The final result for $P_B$ is given by the sum of the weights of all branches that finally reach $\lambda_B$.

\subsection{Rosenbluth}
The Rosenbluth path sampling method is modified by pruning in a similar way to FFS. We focus here only on the Metropolis acceptance/rejection version of the method.  Having generated a point at interface 1 using a free simulation in region $A$, we proceed as described in Section \ref{ros_samp}, until we reach an interface $i$, such that $\lambda_{i-1}> \lambda_A$. We make $k_i$ trial runs from this interface. Each trial run begins with weight $f^{(i)}=1$. As for FFS, trial runs that reach $\lambda_{i-s}$ are terminated with probability $P_p^{i-s}$ and otherwise continue with weight $f^{(i)}$ multiplied by $1/(1-P_p^{i-s})$. After the $k_i$ trials are completed, the ``number of successes'' $N_s^{(i)}$ is defined as the sum of the weights of the trials that eventually reached $\lambda_{i+1}$. This affects the evaluation of the Rosenbluth weight of the partial path up to interface $i$: $W_i = \prod_{j=1}^{i-1} N_s^{(j)}$. This weight is compared with that of the previously accepted partial path up to interface $i$, and, if accepted, $P^{est}_{old}(\lambda_{i+1}|\lambda_i)$ becomes $P^{est}_{new}(\lambda_{i+1}|\lambda_i) = N_s^{(i)}/k_i$. If $N_s^{i}>0$, then one of the successful trials is chosen with probability proportional to its weight $f^{(i)}$. The final point of this path becomes the starting point for shooting trials to the next interface, each of which begins with unit weight $f^{(i+1)}=1$.

\subsection{Test of the pruning algorithms}

\begin{table}[h]
\begin{center}
\begin{tabular}{cccccc}
& $f \times 10^{-1}$ & $P_B \times 10^{-3}$ & $k_{AB} \times 10^{-4}$& $N_{st}\times 10^8$\\
FFS & $ 1.085\pm 0.004 $ & $ 1.38\pm 0.02 $ & $1.50\pm 0.02$  &$4.1$\\
BG  & $1.081\pm 0.004$ & $ 1.36\pm 0.02$ & $  1.47\pm 0.02$ &$2.5$\\
Rb/M  & $ 1.091 \pm 0.003$& $ 1.32\pm 0.02$& $1.44\pm 0.03 $ &$4.1$\\
Rb/WR & $ 1.082 \pm 0.003$& $1.31\pm 0.03$ & $ 1.42\pm 0.03$ &$8.2$\\
\end{tabular}
\end{center}
\caption{FFS and brute force results for
$f={\overline{\Phi}}_{A,0}/{\overline{h}}_{\mathcal{A}}$,
$P(\lambda_n|\lambda_0)$ and $k_{AB}$, for the polymer translocation problem of Section \ref{poltr}, with pruning probability $P_p^{i}=0.5$ at all interfaces.  Units of $f$ and $k_{AB}$ are $D\sigma^{-2}$. Errors represent the standard error in the mean of a series of independent estimates.  $N_{st}$ is the approximate number of simulation steps performed in arriving at the result given in the table. \label{pol_res_pr}}
\end{table}

In order to demonstrate that the pruning procedure described above does lead to correct path sampling, we have repeated the polymer translocation calculations of section \ref{poltr}, using a pruning probability $P_p^{i}=0.5$ for all interfaces. This value for $P_p$ was chosen arbitrarily. All parameters remained the same as those of Section \ref{poltr}: the initial polymer parameter set was used. Table \ref{pol_res_pr} shows the results obtained: on comparison with Table \ref{pol_res}, it is clear that the pruning procedure indeed leads to correct results. Comparing also the total number of simulation steps required to obtain the results of Table \ref{pol_res_pr}, we find that no dramatic improvement in efficiency is achieved by using pruning for this system. For this reason, we did not attempt to optimise $P_p$. Nevertheless, pruning may be of use for other systems.


\end{document}